\documentclass[twocolumn]{aastex631}
\usepackage{showyourwork}
\usepackage{amsmath}
\usepackage{listings}
\usepackage{xcolor}
\usepackage{natbib}

\definecolor{codegreen}{rgb}{0,0.6,0}
\definecolor{codegray}{rgb}{0.5,0.5,0.5}
\definecolor{codepurple}{rgb}{0.58,0,0.82}
\definecolor{backcolour}{rgb}{0.95,0.95,0.92}
\definecolor{mlp}{rgb}{0.294, 0.612, 0.827}

\defcitealias{Haywood2016}{H16}
\defcitealias{Milbourne2019}{M19}
\defcitealias{Ervin2022}{E22}

\graphicspath{{figures/}}

\newcommand{\lt}{<}

\newcommand{\ms}{{\rm m\ s}^{-1}}
\newcommand{\cms}{\ {\rm cm\ s}^{-1}}
\newcommand{\kms}{\ {\rm km\ s}^{-1}}

\newcommand{\Wactive}{W^{\rm active}_{ij}}
\newcommand{\Wquiet}{W^{\rm quiet}_{ij}}
\newcommand{\Wregion}{W^{k}_{ij}}
\newcommand{\Wmu}{W^{\mu}_{ij}}
\newcommand{\PSUAA}{Department of Astronomy \& Astrophysics, 525 Davey Laboratory, The Pennsylvania State University, University Park, PA, 16802, USA}
\newcommand{\PSUCEHW}{Center for Exoplanets and Habitable Worlds, 525 Davey Laboratory, The Pennsylvania State University, University Park, PA, 16802, USA}

\shorttitle{SDO center-to-limb variability}
\shortauthors{M.L. Palumbo III et al.}

\begin{document}
\title{Characterizing Solar Center-to-Limb Radial-Velocity Variability with SDO}

\author[0000-0002-4677-8796]{Michael L. Palumbo III}
\affiliation{\PSUAA}
\affiliation{\PSUCEHW}

\author[0000-0001-7032-8480]{Steven H. Saar}
\affiliation{Center for Astrophysics | Harvard \& Smithsonian, 60 Garden Street, Cambridge, MA 02138, USA}

\author[0000-0001-9140-3574]{Rapha\"{e}lle D. Haywood}
\affiliation{Astrophysics Group, University of Exeter, Exeter EX4 2QL, UK}
\affiliation{STFC Ernest Rutherford Fellow}

\correspondingauthor{Michael L. Palumbo III}
\email{palumbo@psu.edu}

\begin{abstract}
Stellar photospheric inhomogeneities are a significant source of noise which currently precludes the discovery of Earth-mass planets orbiting Sun-like stars with the radial-velocity (RV) method. To complement several previous studies which have used ground- and spaced-based facilities to characterize the RV of the Sun, we here characterize the center-to-limb variability (CLV) of solar RVs arising from various solar-surface inhomogeneities observed by SDO/HMI and SDO/AIA. By using various SDO observables to classify pixels and calculate line-of-sight velocities as a function of pixel classification and limb angle, we show that each identified feature type, including the umbrae and penumbrae of sunspots, quiet-Sun magnetoconvective cells, magnetic network, and plage, exhibit distinct and complex CLV signatures, including a notable limb-angle dependence in the observed suppression of convective blueshift for magnetically active regions. We discuss the observed distributions of velocities by identified region type and limb angle, offer interpretations of the physical phenomena that shape these distributions, and emphasize the need to understand the RV signatures of these regions as astrophysical signals, rather than simple (un)correlated noise processes. 
\end{abstract}

\keywords{radial velocity, solar activity, stellar activity, sunspots}

\section{Introduction} \label{sec:intro}

It has been widely recognized within the astronomical community that magnetically-driven stellar activity poses a significant barrier to the discovery and characterization of Earth-mass planets orbiting solar-like stars with the radial velocity (RV) method \citep{ESS, Decadal, Crass2021}. The sources of RV noise arising from stellar atmospheric phenomena are numerous and complex. Chief among these sources of noise are acoustic oscillations, stochasticity in quiet-Sun magnetoconvection (i.e., granulation and supergranulation), and the alteration of photospheric intensities and velocities by dark sunspots and bright faculae. Even with current- and next-generation RV spectrographs, which boast instrumental precisions at the few tens of$\cms$ level, the detection of Earth-twins (with RV semi-amplitudes of $\sim$10$\cms$) will remain out of reach in the absence of strategies for disentangling these myriad noise sources from the true center-of-mass motion of stars that are incurred by planetary companions. \par 

In response to this challenge, astronomers have turned to the Sun as an essential touchstone. Because the Sun is the only star which we can resolve with high spatial resolution and whose center-of-mass motion relative to the solar system barycenter is precisely and accurately known within $\lesssim 1 \cms$ \citep{Wright2014}, solar observations can be used to understand the impact of intrinsic solar variability on the observed RV variations. Multiple spectrographs, including HARPS-N \citep{Cosentino2014, Dumusque2015}, EXPRES \citep{Jurgenson2016, Llama2022}, and NEID \citep{Schwab2016, Lin2022} have taken advantage of this special knowledge and have been independently monitoring the disk-integrated solar RV with fiber feeds from dedicated solar telescopes. These instruments have enabled astronomers to make headway toward mitigating activity-driven RV variability, but have also demonstrated the magnitude of the problem posed by stellar variability: \citet{AlMoulla2023} calculate that each of oscillations, (super)granulation, and rotation-modulated activity can produce RV signals at a few to several tens of $\cms$ on timescales of minutes to days. \par 

In addition to ground-based spectrographs, space-based observatories that measure velocities on the resolved solar disk have also yielded insight into stellar RV variability, including on much longer timescales. In a seminal study, \citet{Meunier2010a} used MDI/SOHO full-disk Dopplergrams derived from observations of the Ni I $6768$ \AA\ line to reconstruct the disk-integrated solar RV over one solar cycle. Over the full cycle, they report RV variations with amplitude $\sim$8 $\ms$, which are chiefly driven by the suppression of convective blueshift in magnetically active regions. \par

More recently, \citealt{Haywood2016} used HMI/SDO observations of the Fe I $6173$ \AA\ line to construct the disk-integrated solar RV, which they then compared to solar RVs derived from HARPS \citep{Pepe2000} observations of sunlight scattered off the asteroid Vesta. By modeling the total HMI-observed RV as a linear combination of RV contributions from the suppression of convective blueshift and the flux effect (see also \citealt{Dumusque2014}), these authors showed agreement between the HARPS and SDO RVs, validating the use of space-based solar facilities as a tool for understanding solar RV variability, despite their lack of strict RV stabilization as in ground-based fiber-fed spectrographs. \par  

Using the same model for HMI-derived RVs developed in \citet{Haywood2016}, \citealt{Milbourne2019} demonstrated agreement between solar RVs measured directly by the HARPS-N solar telescope. In this same work, they show that overall magnetic filling factor cannot be used alone to completely reconstruct activity-driven RVs; rather, a correlate to active-region size is also necessary. In a subsequent work, \cite{Haywood2022} showed that by modeling RV variations with a linear fit to the disk-averaged unsigned magnetic flux, injected planets with RV semi-amplitudes down to $\sim$0.3 $\ms$ could be recovered. Despite the power of this method, which was able to reduce the RMS of RV variations by 62\%, we currently lack the ability to measure precise disk-averaged magnetic field fluxes for other stars. Together, these works underscore the need to develop informative proxies or new methods of direct measurement for quantifying the magnetic activity of distant stellar photospheres. \par

In addition to RV variability arising from dark and bright photospheric features, stochasticity in quiet-Sun magnetoconvection is a significant and persistent source of RV noise. These convective cells, known as granules, cover the solar surface and introduce a net convective blueshift that changes with viewing geometry \citep{Gray2008}. The amplitude of the convective blueshift is known to vary line-to-line, but generally deeper lines, which have cores that form higher in the solar atmosphere, exhibit lower convective blueshifts \citep{Gray2008, Reiners2016}. Moreover, in disk-resolved light, convective processes create asymmetries in the shapes of lines that vary with center-to-limb angle \citep{Lohner-Bottcher2018b, Stief2019, Lohner-Bottcher2019}. Temporal variations created by the stochastic evolution of convective cells drive variability in both the absolute convective blueshift (i.e., shifts) and asymmetries (i.e., shapes) of lines, constituting another persistent source of RV noise. \par 

In order to characterize the aggregate effect of these magnetoconvective contributions to the solar RV, \citet{Cegla2018, Cegla2019} used magnetohydrodynamic (MHD) simulations of solar convective cells to model the disk-integrated RV of the quiet photosphere. In \citet{Cegla2018}, they show that tiling a model stellar grid with their simulated patches (consisting of granules, intergranular lanes, and magnetic bright points) can reproduce the characteristic convective blueshift center-to-limb variability (CLV) observed in solar lines \citep[e.g.,][]{Cavallini1985, Lohner-Bottcher2018b, Lohner-Bottcher2019}. Further applying this parameterization in \citet{Cegla2019}, they showed that diagnostics of line asymmetry (e.g., the bisector inverse slope) can be used to remove a significant fraction of the RV ``noise" created by magnetoconvective motions. \par 

The findings presented in \citet{Cegla2018} and \citet{Cegla2019} emphasize that stellar RV ``noise'' is in reality the composite of various signals that arise from complex physical phenomenon, which must be understood as such in order to achieve true $10 \cms$ wholesale Doppler precision. Following the line-shape analysis presented by \citet{Cegla2019}, \citet{Sulis2023} attempted to connect variations in the shapes of cross-correlation functions (CCFs) to changes in RVs for two relatively Sun-like stars observed by ESPRESSO \citep{Pepe2021}. After binning the observations to average out the stellar $p$-modes, only hints of a correlation between the CCF curvature and the measured RVs were retrieved for one star, and no significant correlation was found for the other. Commenting on this result, the authors of \citet{Sulis2023} noted that their analysis was complicated by difficulties in filtering the $p$-modes, as well as SNR limitations, since correlations between RV and shape of the sort found by \citet{Cegla2019} are expected to be strongly limited by photon noise. In their conclusion, the authors of \citet{Sulis2023} emphasized that future observations will be needed to understand the precise impact of granulation on CCF curvature and RV shifts. \par 

Pending such future observational studies of other stars, complementary solar observations provide the best opportunity for characterizing the impacts of magnetoconvection (and the suppression thereof) on individual lines. In this work, we turn back to the SDO data in order to provide an observational characterization of the center-to-limb dependence of RV variability arising from various inhomogeneities in the solar atmosphere as probed by the Fe I 6173 \AA\ line observed by HMI. The inhomogeneities we probe in this work include spots (both penumbrae and umbrae), the quiet Sun, magnetic network, and chromospheric plage/photospheric faculae. In \S\ref{sec:data}, we describe our initial processing of the SDO data used in this work and summarize the computation of RVs from SDO/HMI data. In \S\ref{sec:region_id}, we detail our process for the classification of solar surface features from SDO data, noting the differences between our schema and those used in \citet{Yeo2013}, \citet{Haywood2016}, \citet{Milbourne2019}, and \citealt{Ervin2022}. In \S\ref{sec:clv}, we present our computed RVs and discuss the different center-to-limb variability seen for each identified region type, particularly noting the limb-angle dependence of the suppression of convective blueshift observed in spots, plage, and magnetic network. In \S\ref{sec:discussion}, we offer physical interpretations of the velocity distributions seen in certain regions, and contextualize the implications of these findings for the pursuit of 10$\cms$ RV precision and the search for extrasolar Earth-twins. \par

\section{SDO Data Sampling and Processing} \label{sec:data}

\begin{figure*}[!htb]
    \gridline{\fig{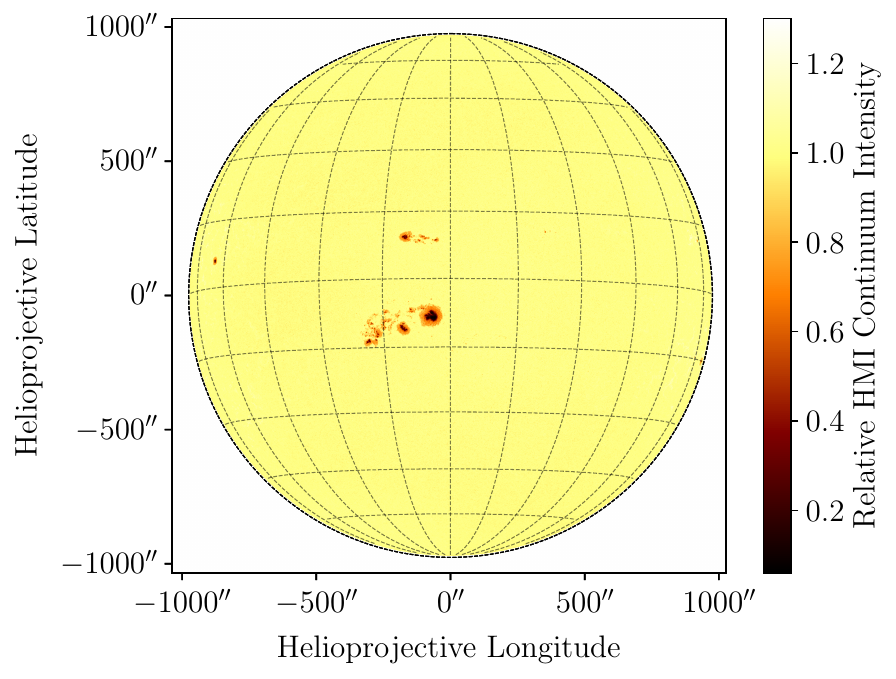}{0.46\textwidth}{(a)}
              \fig{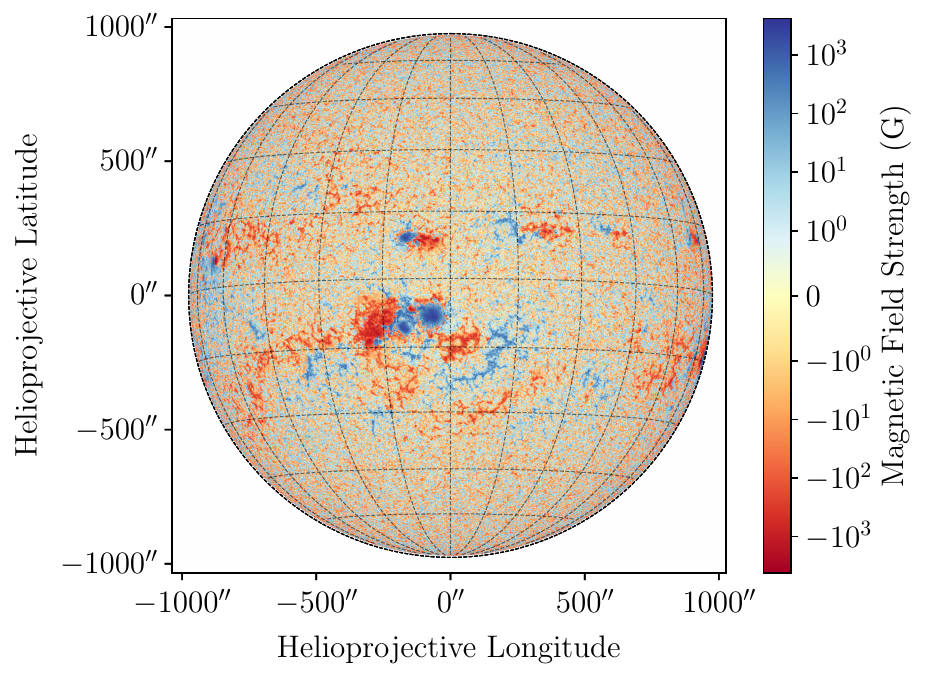}{0.48\textwidth}{(b)}}
    \gridline{\fig{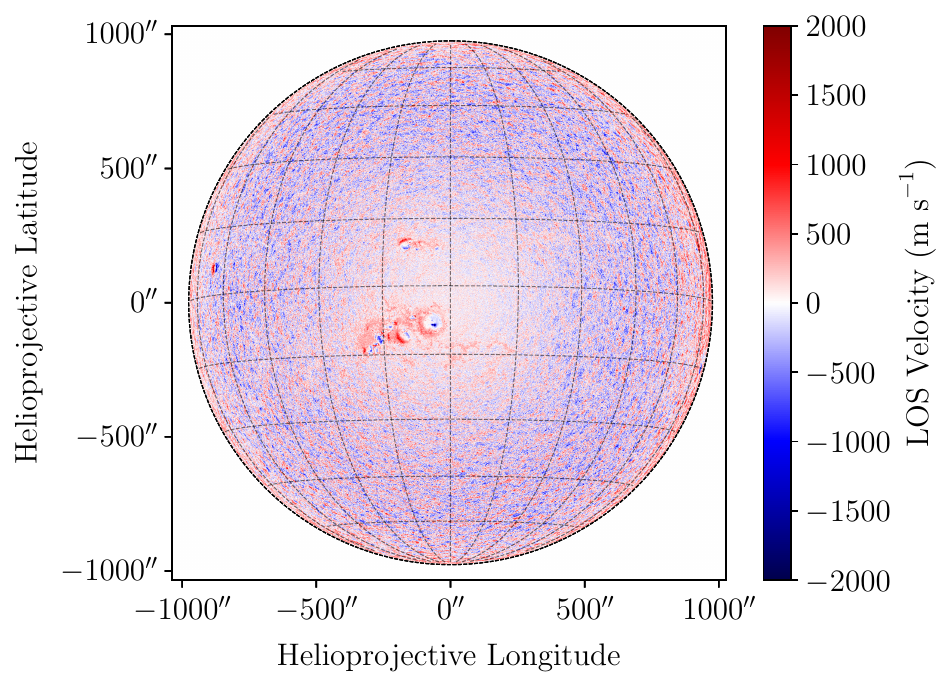}{0.48\textwidth}{(c)}
              \fig{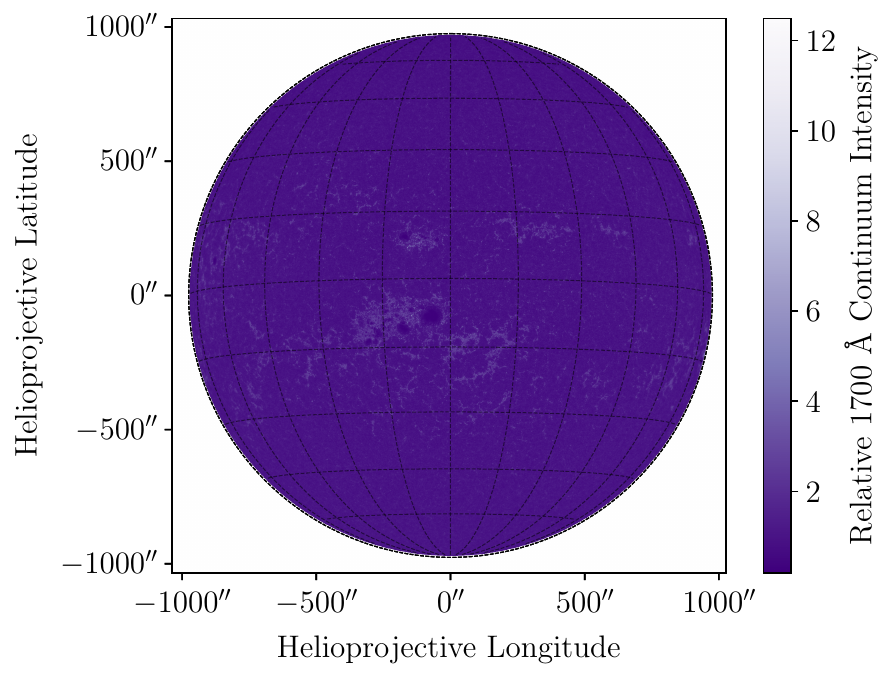}{0.46\textwidth}{(d)}}
    \caption{Example processed SDO/HMI and SDO/AIA data observed on 2014 January 07, near the peak of Solar Cycle 24. Note the difference in appearances in the sunspot group near disk center between panels. As with all other calculations and analyses in this work, pixels with $\mu < 0.1$ are masked. \textit{Panel (a)}: SDO/HMI continuum filtergram corrected for limb darkening. \textit{Panel (b)}: SDO/HMI line-of-sight magnetogram corrected for foreshortening. \textit{Panel (c)}: SDO/HMI Doppplergram corrected for gravitational redshift, solar differential rotation, meridional circulation, and relative spacecraft motion using the procedure presented in \S\ref{subsec:processing}. Large velocity perturbations are readily apparent in the penumbrae and umbrae of the active regions, the effects of which are discussed in \S\ref{subsec:spot_vels}. \textit{Panel (d)}: SDO/AIA 1700 \AA\ continuum filtergram reprojected onto the HMI plate scale and corrected for limb brightening. As discussed in \S\ref{subsec:bright_id}, we use the 1700 \AA\ continuum images to supplement our identification of plage and magnetic network, which tend to have lower contrast near disk center in SDO/HMI continuum filtergrams (see \S\ref{subsec:bright_id}).}
    \label{fig:one}
    \script{fig1.py}
\end{figure*}

In order to characterize the center-to-limb dependence of radial velocity perturbations introduced by different magnetic features, we use data from the Solar Dynamics Observatory's (SDO) Helioseismic Magnetic Imager \citep[HMI;][]{Scherrer2012} and Atmospheric Imaging Assembly \citep[AIA;][]{Lemen2012}. From SDO/HMI (hereafter HMI), we use the 720-second Level 1.5 continuum filtergrams, line-of-sight Dopplergrams, and line-of-sight magnetograms. The reduction processes for these data products are described in \citet{Couvidat2016}. From SDO/AIA (hereafter AIA), we use Level 1.5 1700 \AA\ continuum filtergrams. The reduction process for AIA data products is described in \citet{Lemen2012}. Following our initial data processing and cleaning (\S\ref{subsec:processing}), we use the data to robustly identify different region types (\S\ref{sec:region_id}), and then reconstruct the relative contribution of these features to the total ``Sun-as-a-star" radial velocity as a function of limb position (\S\ref{sec:clv}). \par

The Python source code used to process the SDO data is available on GitHub\footnote{\url{https://github.com/palumbom/sdo-clv-pipeline}}, and version-tagged releases are archived on Zenodo \citep{palumbo_michael_louis_2023_8273623}. Data products used to generate figures and other results in this work are also available from Zenodo \citep{palumbo_iii_michael_louis_2023_8273650}. \par 

\subsection{Data Sampling} \label{subsec:downloading}

We obtain the SDO data used in this work from the Joint Solar Operations Center (JSOC) archive through the \texttt{SunPy} JSOC client \citep{SunPyCommunity2020}. We sample every fourth hour for each calendar day in the years 2012-2015, capturing the years around the peak of solar activity during Solar Cycle 24 \citep[][]{sidc}. These six daily samples provide sufficient temporal resolution such that any equatorial patch of the solar surface at the spatial resolution of HMI falls within every $\mu$ bin along its rotation path. As explained in \S2.1.2 of \citet{Haywood2022}, these multiple daily samples also allow us to average over the uncertainty in the spacecraft velocity, which manifests as a sinusoidal shift in the Dopplergrams with periodicity of 12 and 24 hours (see also the discussion of the Dopplergram systematics in Appendix~\ref{app:vel_components}).  \par 

We exclude any observations conducted during spacecraft maneuvers, eclipses by the Earth and Moon, and the 2012 transit of Venus\footnote{\url{https://aia.lmsal.com/public/sdo_spacecraft_night.txt}}, as well as any observations with data quality issues flagged in the FITS header (237 in total). In a small number of cases (24), observations were excluded due to errors in the flattening of the filtergrams (\S\ref{subsubsec:limb_dark}) or the correction of the Dopplergram (\S\ref{subsubsec:dopplergram_correct}). In total, the data and analyses presented in this work are based on 8312 observation epochs. \par 

\subsection{Initial Data Processing} \label{subsec:processing}

In order to use SDO data to identify magnetic regions and compute their contributions to the total solar disk-integrated radial velocity, we first apply several corrections to the HMI and AIA data. This processing is largely carried out in a manner similar to that described in \citet{Haywood2016} and \citet{Ervin2022}, with some exceptions which we highlight in this section. Example processed HMI and AIA are shown in Figure~\ref{fig:one}. \par

\subsubsection{Coordinate Transformations, Data Interpolation, and Foreshortening Corrections}

As a first step, we perform a change of coordinate systems from the Helioprojective Cartesian frame to the Heliographic Carrington frame using the information contained in the FITS headers in each of the HMI and AIA images. The mathematical formalism describing this transformation is laid out in \citet{Thompson2006}, and the implementation thereof is described in detail in \citet{Haywood2016} and \citet{Ervin2022}. \par 

Following the coordinate transformations, we account for differences in the plate scales of HMI and AIA by aligning and resampling the AIA 1700 \AA\  continuum images to the corresponding HMI continuum filtergrams using the \texttt{reproject\_interp} function provided by \texttt{astropy}\footnote{\url{https://reproject.readthedocs.io/en/stable/}} \citep{AstropyCollaboration2013, AstropyCollaboration2018, AstropyCollaboration2022}. We use the default bilinear interpolation for this reprojection. \par 

Following these coordinate transformations and resamplings, we perform a number of additional corrections specific to each SDO data product. To account for foreshortening in the HMI magnetograms, we approximate the true magnetic field strength in each pixel $B_{r, ij}$ in the same manner as \citet{Haywood2016} and \citet{Ervin2022}:

\begin{equation}
    B_{r,ij} = B_{{\rm obs}, ij}/ \mu_{ij}
\end{equation}

\noindent where  $B_{{\rm obs}, ij}$ is the observed magnetic field strength in each pixel, and $\mu_{ij}$ is the cosine of the angle $\theta_{ij}$ subtended by the solar surface vertical and the line of sight at pixel $ij$, i.e., $\mu_{ij} = \cos(\theta_{ij})$. The foreshortening-corrected magnetogram is also used to identify active and quiet pixels (see \S\ref{sec:region_id}). We use the same threshold defined in \citet{Yeo2013} and used by \citet{Haywood2016} and \citet{Ervin2022}:

\begin{equation} \label{eq:mag_thresh}
    B_{r,\ {\rm thresh}} = \left| 24\ {\rm G}/\mu_{ij}\right|
\end{equation}

\noindent As in \citet{Haywood2016}, we do not classify any isolated pixels with $|B_{r, ij}| > B_{r {\rm thresh}}$ as active. \par 

\subsubsection{Dopplergram Corrections} \label{subsubsec:dopplergram_correct}

Our treatment of HMI Dopplergrams differs somewhat from the procedure described in \citet{Haywood2016} and \citet{Ervin2022}. In addition to correcting for the Doppler shifts introduced by solar differential rotation ($v_{ij,\ {\rm rot}}$) and relative spacecraft motion ($v_{ij,\ {\rm sat}}$), we also explicitly calculate and correct for additional shifts arising from meridional circulation ($v_{ij,\  {\rm mer}}$), as well as constant offsets for the gravitational redshift ($v_{\rm grav}$) and the convective blueshift ($v_{\rm cbs}$). Following the calculation of these velocities, the corrected Dopplergram is computed as:

\begin{equation} \label{eq:v_corr}
\begin{split}
    v_{ij,\ {\rm corr}} =\ v_{ij} & - v_{\rm grav}\ - v_{ij,\  {\rm mer}}\ - \\
                          & - v_{ij,\ {\rm sat}} - v_{ij,\ {\rm rot}}\ - \\
                          & - v_{\rm cbs}
\end{split}
\end{equation}

\noindent Examples for each of these velocity components are plotted in Figure~\ref{fig:vel_components} in Appendix~\ref{app:vel_components}, and the calculation of each component is discussed below. \par

As in \citet{Haywood2016} and \citet{Ervin2022}, we obtain $v_{ij,\ {\rm sat}}$ by projecting the spacecraft velocity components from the FITS header onto the line-of-sight vector for each pixel. For $v_{\rm grav}$, we assume a constant shift of $633\ \ms$, corresponding to the gravitational redshift from the solar photosphere to 1 A.U., as measured by \citet{GonzalezHernandez2020}. We calculate the velocities from differential rotation ($v_{ij,\ {\rm rot}}$) and meridional circulation ($v_{ij,\ {\rm mer}}$) following \citet{Kashyap2021}\footnote{See also \url{https://github.com/samarth-kashyap/hmi-clean-ls}}, who perform a least-squares fit of the odd-degree toroidal components and even-degree poloidal components, respectively, of the vector spherical harmonic expansion to the HMI Dopplergram data (see their Section 2 and Appendix A). Notably \citet{Kashyap2021} explicitly fit for a fifth-order, limb-angle dependent convective blueshift ($v_{\rm cbs}$) as:

\begin{equation}
    v_{\rm cbs}(\mu_{ij}) = \sum_{n=0}^5 c_n P_n(\mu_{ij}){\rm ,}
\end{equation}

\noindent where $c_n$ is the $n^{\rm th}$ coefficient and $P_n(\mu_{ij})$ are the shifted Legendre polynomials defined in Appendix B of \citet{Kashyap2021}. Whereas \citet{Kashyap2021} sought to correct for the convective blueshift limb-angle trend in their analysis, we instead need to retain this trend in order to accurately calculate the suppression of convective blueshift (see Equation~\ref{eq:vconv}). Consequently, we fit only for the zeroth-order coefficient (corresponding to $P_0(\mu_{ij}) = 1$) yielding a $v_{\rm cbs}$ that is constant with $\mu$. \par 

After applying these corrections to the raw Dopplergram data, the mean of the resulting $v_{ij,\ {\rm corr}}$ map (which is distinct from the intensity-weighted mean in Equations~\ref{eq:vhat} and \ref{eq:vquiet}) is approximately zero for observation epochs without significant magnetic active regions on the disk. An example Dopplergram corrected by this procedure is shown in Panel (c) of Figure~\ref{fig:one}, as well as in the bottom-right panel of Figure~\ref{fig:vel_components} in Appendix~\ref{app:vel_components}. \par 

\subsubsection{Limb Darkening and Limb Brightening} \label{subsubsec:limb_dark}

In our processing of the continuum filtergrams, we model and remove the effects of limb darkening (in the case of HMI filtergrams) or limb brightening (in AIA 1700 \AA\ filtergrams). Rather than using prescribed limb darkening coefficients, as in \citet{Haywood2016} and \citet{Ervin2022}, we instead fit each filtergram with a second-order limb darkening/brightening relation of the form presented by \cite{Kopal1950}:

\begin{equation} \label{eq:limb_dark}
    \frac{I(\mu)}{I_0} = 1 - c_1 (1- \mu) - c_2(1-\mu)^2
\end{equation}

\noindent where $I_0$ is a scaling factor such that $I(\mu = 1) = 1$ for a theoretical uniform, featureless disk. By fitting for  $I_0$ and the coefficients $c_1$ and $c_2$ for each observation, we ensure that the intensity normalization does not vary between observation epochs, due to e.g., detector changes over the course of the mission. To account for variations introduced by surface features (e.g., faculae, spots), we divide the solar disk into many annuli with bin centers linear in $\mu$ and fit to the average brightness within each annulus, excluding pixels with intensities that are more than $2\sigma$ removed from the mean intensity of the annulus. We then compute the flattened continuum filtergram as:

\begin{equation}
    I^{\rm flat}_{ij} = \frac{I_{ij}}{L_{ij}},
\end{equation}

\noindent where $L_{ij}$ is the best-fit limb darkening/brightening model evaluated at each pixel. \par 

Because of data quality issues, introduced by both the extreme foreshortening near the solar limb and deviations from the simple limb-darkening law used, we exclude all pixels with $\mu_{ij} < 0.1$, such that they are not included in any velocity calculations or region identifications described in the following sections. Because these pixels constitute $\sim$1$\%$ of the total on-disk pixels observed by HMI and contribute $\lesssim$0.5$\%$ of the total disk-integrated HMI continuum light, the exclusion of these pixels does not significantly affect the results presented in this work. \par

\subsection{Calculation of Sun-as-a-star Velocities} \label{subsec:sun_vels}

\begin{figure*}[!htb]
    \epsscale{1.1}
    \gridline{\fig{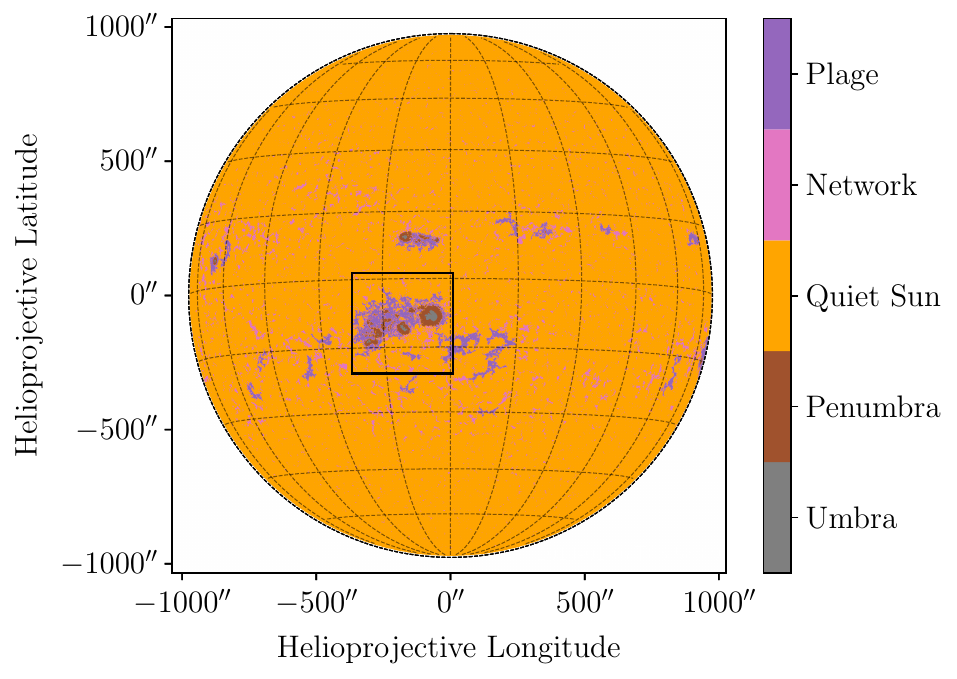}{0.52\textwidth}{}
              \fig{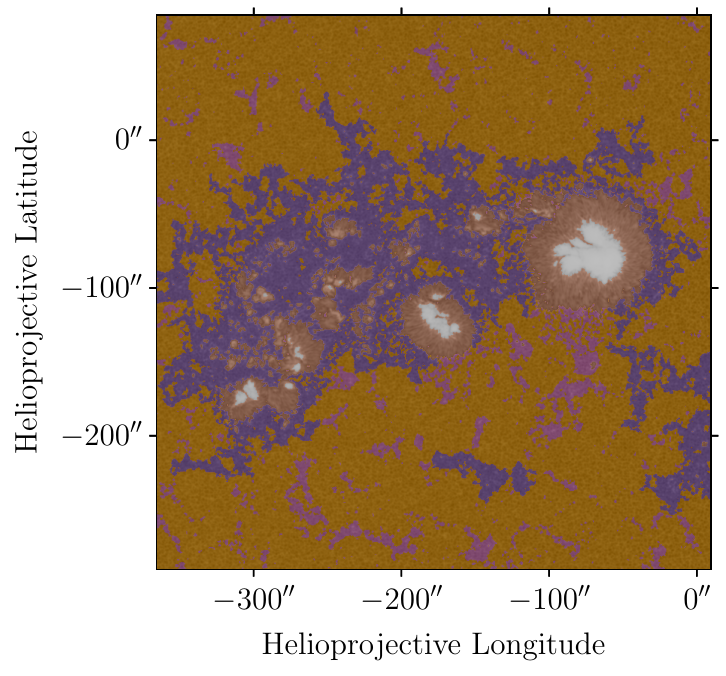}{0.40\textwidth}{}}
    \caption{\textit{Left:} Example region identification from the processed data shown in Figure~\ref{fig:one}. Pixels are color-coded by classification. The same color scheme is used for curves in subsequent figures in this text. \textit{Right:} Zoom-in on the large active region near disk center, corresponding to the region bounded by the black box in the plot at left. The classification mask (same color scale as at left) is overplotted with transparency on the flattened HMI continuum (grayscale) to demonstrate faithful identification of region types. Umbrae and penumbrae are well-separated by our identification scheme, and the surrounding plage/faculae and magnetic network are successfully identified despite their low contrast with quiet-Sun pixels in the HMI continuum near disk center (see \S\ref{subsec:bright_id}).}
    \label{fig:two}
    \script{fig2.py}
\end{figure*}

Using HMI data, \citet{Haywood2016} calculate whole-Sun RVs ($\hat{v}$), as well as RV contributions from the quiet Sun ($\hat{v}_{\rm quiet}$), the photometric effect of dark and bright regions ($\Delta \hat{v}_{\rm phot}$), and the suppression of convective blueshift ($\Delta \hat{v}_{\rm conv}$). We follow the modified procedure for calculating these velocities described in \citet{Milbourne2019} and \citet{Ervin2022}, with the addition of weights used to calculate these terms for specific region classifications (e.g., umbrae, penumbrae, plage, etc.) identified using the procedures outlined in \S\ref{sec:region_id}. \par   

Adapting the equation for the disk-integrated RV from \citet{Haywood2016}, \citet{Milbourne2019}, and \citet{Ervin2022}, we calculate the RV for a given region classification $k$ within a given $\mu$ bin, $\hat{v}_{k, \mu}$, as the intensity-weighted mean of the corrected Dopplergram velocities ($v_{ij, \rm{corr}}$ - as given by Equation~\ref{eq:v_corr}):

\begin{equation} \label{eq:vhat}
    \hat{v}_{k, \mu}=\frac{\sum_{i j}v_{ij, {\rm corr}} I_{i j} \Wregion \Wmu}{\sum_{i j} I_{i j} \Wregion \Wmu}
\end{equation}

\noindent where $\Wregion$ and $\Wmu$ are additional binary weights used to select the region classification and $\mu$ range of interest, respectively. For example, to compute $\hat{v}_{k, \mu}$ for plage in the range $0.9 < \mu \leq 1.0$, $\Wregion$ would be set to $1$ for all pixels classified as plage (and 0 elsewhere), and $\Wmu$ would be set to $1$ for all pixels in the previously specified $\mu$ range (and 0 elsewhere). In the limiting case where all elements of $\Wregion$ and $\Wmu$ are $1$, Equation~\ref{eq:vhat} becomes equivalent to Equation 27 of \citet{Haywood2016} for the disk-integrated $\hat{v}$ (modulo the differences in our computation of $v_{ij, {\rm corr}}$). \par 

We calculate the RV of the quiet Sun as in the previous works, but again with the inclusion of the additional weight terms:

\begin{equation} \label{eq:vquiet}
    \hat{v}_{{\rm quiet}, \mu}=\frac{\sum_{i j}v_{ij, {\rm corr}} I_{i j} \Wquiet \Wmu}{\sum_{i j} I_{i j} \Wquiet \Wmu}
\end{equation}

\noindent where $\Wquiet$ is set to 1 for pixels identified as quiet Sun, and 0 otherwise. By definition, $\hat{v}_{k, \mu}$ evaluated for the quiet Sun is equal to $\hat{v}_{{\rm quiet}, \mu}$, and $\hat{v}_{{\rm quiet}, \mu} = 0$ for umbrae, penumbrae, network, and plage. \par

Following from the \citet{Milbourne2019} and \citet{Ervin2022} expression for the suppression of convective blueshift, we compute $\Delta \hat{v}_{\text {conv}, k, \mu}$ as the difference: 

\begin{equation} \label{eq:vconv}
    \Delta \hat{v}_{\text {conv}, k, \mu} = \hat{v}_{k, \mu}-\hat{v}_{{\rm quiet}, \mu}
\end{equation}

\noindent By definition, $\Delta \hat{v}_{\text {conv}, k, \mu} = 0 $ for the quiet Sun. In the case where $\hat{v}_{k, \mu} = 0$ (i.e., no pixels were classified as the given region type within that $\mu$ bin) , we set $\Delta \hat{v}_{\text {conv}, k, \mu}$ = 0 and exclude these instances from the calculations of the means and distributions presented in \S\ref{sec:clv} and the figures therein. \par 

\subsubsection{Choice of Weighting Scheme}

We note that the our choice of weighting scheme normalizes the computed velocities to the sum of intensities for a given region classification and $\mu$ bin. This ``localized" normalization removes the effects of limb darkening, allowing us to compare velocities for a given region classification across the disk (see \S\ref{sec:clv}). However, as a consequence of this choice, one cannot simply sum over all computed velocities $\hat{v}_{k, \mu}$ to obtain the disk-integrated velocity $\hat{v}$ as defined in \citet{Haywood2016}:

\begin{equation}
    \hat{v} \neq \sum_{k, \mu} \hat{v}_{k, \mu} 
\end{equation}

\noindent Rather, we must weight each $\hat{v}_{k, \mu}$ by the fraction of the total disk-integrated intensity emitted by the region, $f_{k, \mu}$:

\begin{equation}
    \hat{v} = \sum_{k, \mu} f_{k, \mu} \hat{v}_{k, \mu},
\end{equation}

\noindent where $f_{k, \mu}$ is given by:

\begin{equation}
    f_{k, \mu} = \frac{\sum_{ij} I_{ij} \Wregion \Wmu}{\sum_{ij}\Wregion \Wmu}
\end{equation}

\noindent By definition, the sum over all $f_{k, \mu}$ is 1. 

\section{Region Identification} \label{sec:region_id}

In order to explore the center-to-limb dependence of the solar radial velocity contribution, we use various thresholds to classify pixels into different surface-feature categories. Relative to past works which classified SDO pixels \citep[e.g.,][etc.]{Yeo2013, Milbourne2019}, we use an expanded classification scheme consisting of umbrae, penumbrae, quiet Sun, magnetic network, and plage in order to more finely assess the RV impact of various solar surface features. We plot the solar disk with example pixel classification in Figure~\ref{fig:two}. \par 

To distinguish umbrae, penumbrae, and quiet Sun pixels (\S\ref{subsec:spot_id}), we implement intensity thresholds analogous to those first described by \citet{Yeo2013} and subsequently used by \citet{Haywood2016} and \citet{Ervin2022}. To more robustly identify and separate magnetic network and plage (\S\ref{subsec:bright_id}), we supplement the identification procedure carried out by \citet{Milbourne2019} with additional thresholds derived from AIA $1700$ \AA\ continuum filtergrams. As in these prior works, we do not classify pixels near the solar limb ($\mu_{ij} \lt 0.1$) where the data quality becomes spurious and foreshortening extreme. \par 

\subsection{Identifying Penumbrae and Umbrae} \label{subsec:spot_id}

\begin{figure*}[!htb]
    \epsscale{1.1}
    \plottwo{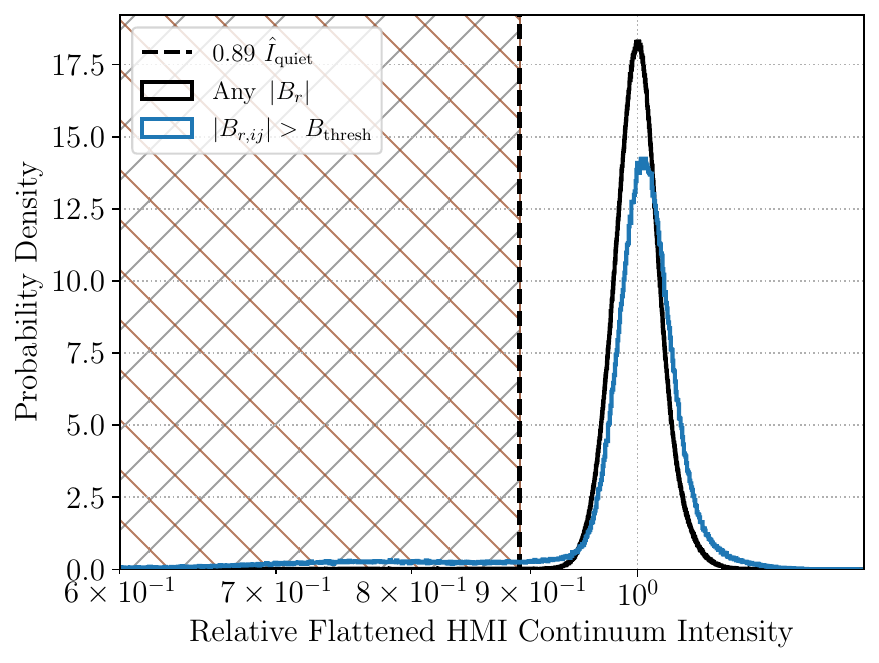}{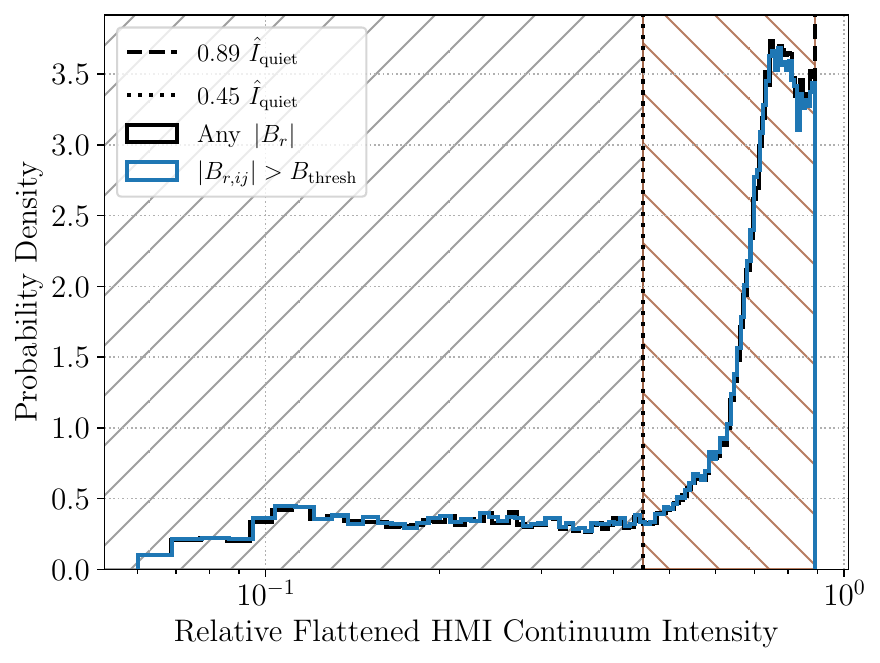}
    \caption{Distributions of flattened and normalized HMI continuum intensities for all pixels (black curve) and magnetically active pixels ($\left| B_{r, ij}\right| > B_{\rm thresh}$, blue curve) in a 720-second HMI continuum image observed on 2014 January 07. \textit{Left:} Full distribution of HMI continuum intensities. The dashed vertical line indicates the threshold used by \cite{Yeo2013} and \citet{Haywood2016} to distinguish spots (brown and gray cross-hatched region) from the quiet Sun (in the case of magnetically inactive pixels) and faculae (in the case of magnetically active pixels). \textit{Right:} Distribution of HMI continuum intensities below $0.89\ \hat{I}_{\rm quiet}$, characterized by a long flat tail and a rising slope which peaks just below $0.89\ \hat{I}_{\rm quiet}$. We find that using the location of the onset of the sloped region (about $0.45\ \hat{I}_{\rm quiet}$) reliably separates darker umbrae (gray forward-hatched region) from lighter penumbrae (brown back-hatched region).}
    \label{fig:three}
    \script{fig3.py}
\end{figure*}

\citet{Yeo2013} showed that an intensity threshold can robustly separate quiet Sun pixels from sunspots and pores:

\begin{equation}
    I^{\rm QS}_{\rm thresh} = 0.89\ \hat{I}_{\rm quiet},
\end{equation}

\noindent where $\hat{I}_{\rm quiet}$ is the mean flattened intensity of magnetically-inactive pixels (i.e., $|B_{r, ij}| \lt B_{\rm thresh}$). This threshold is shown in the left-hand panel of Figure~\ref{fig:three}. Pixels with continuum intensities less than that demarcated by the vertical dashed black line are classified as spots by \citet{Yeo2013}.\par 

To further classify the identified spots into penumbrae and umbrae regions, we examined the distribution of HMI continuum intensities below $I^{\rm QS}_{\rm thresh}$. As seen at right in Figure~\ref{fig:three}, the distribution of these intensities exhibits a flat region at low intensity, which then begins to rise into a peak just below $0.89\ \hat{I}_{\rm quiet}$. By using the approximate boundary between these regions as an additional intensity threshold, we are able to satisfactorily separate out umbrae and penumbrae. Extending the methodology of \cite{Yeo2013}, a threshold of

\begin{equation}
    I^{\rm spot}_{\rm thresh} = 0.45\ \hat{I}_{\rm quiet},
\end{equation}

\noindent corresponds to the location where the gradient of pixel intensities begins to rise into the local peak below $I^{\rm QS}_{\rm thresh}$. Pixels with $I^{\rm flat}_{ij} < I^{\rm spot}_{\rm thresh}$ are designated umbrae, and pixels with $I^{\rm spot}_{\rm thresh} < I^{\rm flat}_{ij} < I^{\rm QS}_{\rm thresh}$ are designated penumbrae. \par

\subsection{Identifying Plage and Magnetic Network} \label{subsec:bright_id} 

\begin{figure*}
    \epsscale{1.15}
    \plotone{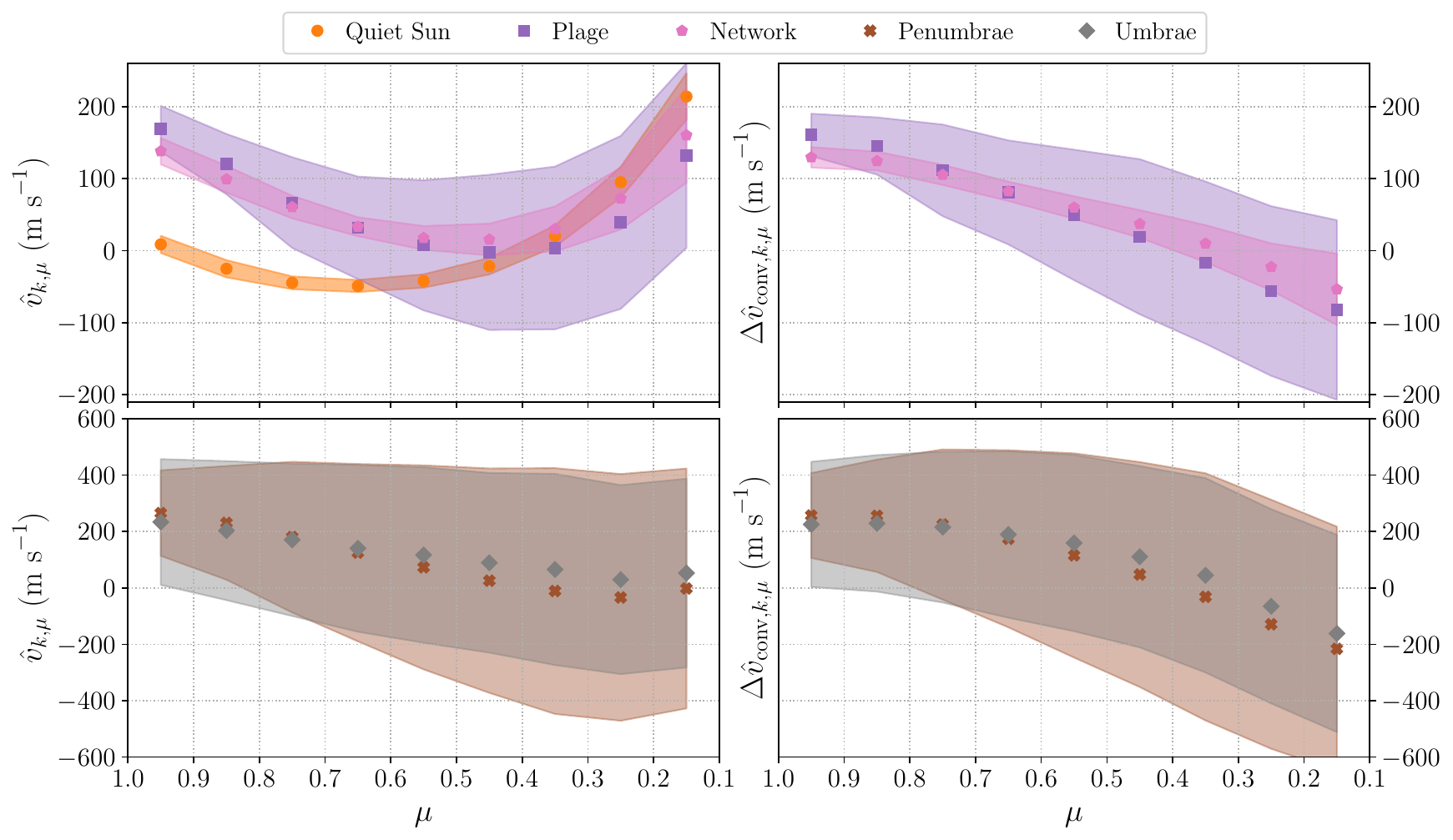}
    \caption{Identified region types exhibit distinct center-to-limb trends in total velocity $\hat{v}$ (left-hand column) and $\Delta \hat{v}_{\rm conv}$ (right-hand column). Note the reversed $x$-axis, with disk-center ($\mu = 1$) at left and the limb ($\mu = 0$). Points indicate the mean velocity in each $\mu$ bin. Shaded regions indicate the 1$\sigma$ width of the velocity distributions at each point. Penumbrae and umbrae (bottom panels) are plotted separately from the other region types (top panels) due to the large differences in the velocity scales for these regions. By definition, $\Delta \hat{v}_{\rm conv}$ is 0 for the quiet Sun (orange circles), so no distribution is shown for this region type in the right-hand column. The distinctive angular dependence of the suppression of convective blueshift is apparent in the right-hand panels.}
    \label{fig:clv}
    \script{fig4_5_6.py}
\end{figure*}

\cite{Yeo2013} and \citet{Haywood2016} designated all pixels with $|B_{r,\ ij}| > B_{\rm thresh}$ and $I_{{\rm flat},\ ij} > I_{\rm thresh,\ QS}$ as faculae. However, this classification scheme does not account for varying contrast between faculae and quiet Sun as a function of limb angle. In the ``bright'' or ``hot wall'' model first discussed by \citet{Spruit1976} and verified via MHD simulations by \citet{Keller2004}, faculae viewed in the optical continuum (e.g., the HMI 6173 \AA\ continuum) appear brighter near the limb because the line-of-sight intersects a longer path length in the thin, brightness-enhanced region atop integranular lanes (see Figure 4 of \citealt{Keller2004}). Because the faculae contrast is reduced near disk-center, it is possible that areas of slightly enhanced magnetic fields (and therefore altered velocity flows) could be misclassified as quiet Sun using information from HMI alone. To address this issue, we use AIA 1700 \AA\ continuum intensity maps to bolster our classification of these bright regions. The 1700 \AA\ continuum, which probes the middle-to-upper photosphere near the temperature minimum \citep{Fossum2005}, exhibits enhanced contrast between regions of bundled magnetic field lines (i.e., network, plage) across the disk \citep{Yeo2019}. \par 

To define an intensity threshold for identifying plage and network in the AIA 1700 \AA\ continuum images, we adapt the procedure used in \cite{Yeo2013} to define their faculae intensity threshold for the HMI 6173\ \AA\ continuum. We find that a threshold calculated as 

\begin{equation}
    I^{\rm AIA}_{\rm thresh} = \frac{\Sigma_{ij}I^{\rm AIA}_{\rm flat}\Wactive }{\Sigma_{ij}\Wactive }
\end{equation}

\noindent cleanly identifies additional magnetic bright regions that exhibit lower contrast in the HMI continuum. Because we use an additional criterion derived from the 1700 \AA\ continuum images to identify these bright regions, we refer to them as ``plage'' and ``network,'' rather than ``faculae,'' following the nomenclature of \citet{Milbourne2019}, which is also consistent with the convention of referring to these regions as ``plage'' when seen in light originating above the photosphere. To avoid biasing the threshold, we exclude any pixels already flagged as penumbrae or umbrae from this calculation, which appear dark in the 1700 \AA\ filtergrams, and would artificially lower the intensity threshold. To prevent any false positive introduced by detector noise or other systematics, we additionally exclude any isolated bright pixels from this classification criterion. \par 

To explore and analyze any potential differences in the radial velocities and center-to-limb behavior of smaller and larger magnetic bright regions, we further discriminate these pixels into magnetic network and plage classifications. Whereas faculae are generally concentrated into dense regions and contiguous with chromospheric plage, regions of magnetic network are sparser and more elongated \citep[see the introduction of][for a review]{Buehler2019}. In order to classify these regions in HMI observations, \citet{Milbourne2019} showed that there is a sharp cutoff in the 2D distribution of active region co-latitude and area at about $\sim$20 microhemispheres (or equivalently $\sim$60 Mm$^2$; see their Figure 6 and \S4.4). Adopting this criterion, we set:

\begin{equation}
    A_{\rm thresh} = 20\ {\rm microhemispheres}
\end{equation}

\noindent I.e., plage are bright, magnetically active regions with areas greater than 20 parts per million (ppm) of the visible solar hemisphere (quantified by pixel count), and magnetic network with areas lesser than this threshold. \par

To measure region areas, we first use \texttt{scipy.ndimage.label} \citep{2020SciPy-NMeth} to identify and assign a unique integer label to all contiguous regions meeting the intensity thresholds defined for HMI and AIA. Diagonally-adjacent pixels are counted as contiguous in our labeling scheme. With the separate regions identified and labeled, we use \texttt{skimage.measure.regionprops} \citep{van2014scikit} to compute the area in pixels of each discrete region. Contiguous regions with areas larger than or equal to $A_{\rm thresh}$ are then designated as plage, and all smaller regions as magnetic network. \par 

\section{Center-to-Limb RV Dependence} \label{sec:clv}

\begin{figure*}
    \epsscale{1.15}
    \plotone{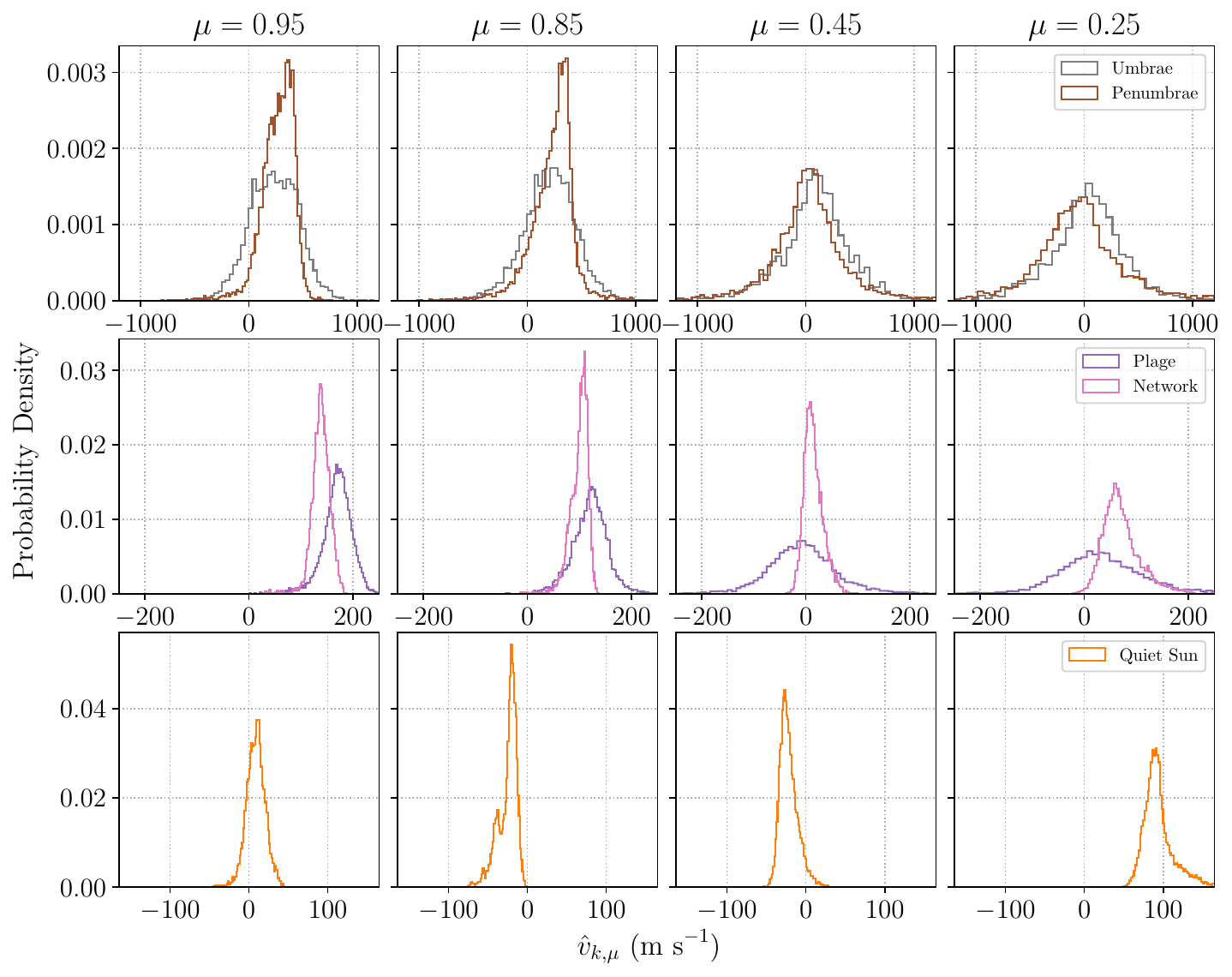}
    \caption{Distribution of $\hat{v}$ by region type (rows) for a subset of $\mu$ bins (columns). The $\mu$ value labeled at the top of each row denotes the center of each bin considered. Note the different $x$- and $y$-axis scales between rows. \textit{Top row:} Penumbrae distributions evolve distinctly in shape with limb angle compared to the umbrae distributions. Near disk-center, the penumbrae distributions are narrower and more peaked than the umbrae distributions. Closer to the limb, the penumbrae distribution is comparatively broad and more closely follows the distribution of umbrae velocities. \textit{Middle row:} Across limb angles, the distribution of network velocities are comparatively peaked compared to the plage velocity distributions, especially near the limb. \textit{Bottom row:} The distribution of quiet Sun velocities has a noticeable redshifted tail near the limb, potentially corresponding to plasma at the top of convective cells with velocity tangent to the solar surface.}
    \label{fig:hist}
    \script{fig4_5_6.py}
\end{figure*}

\begin{figure*}
    \epsscale{1.15}
    \plotone{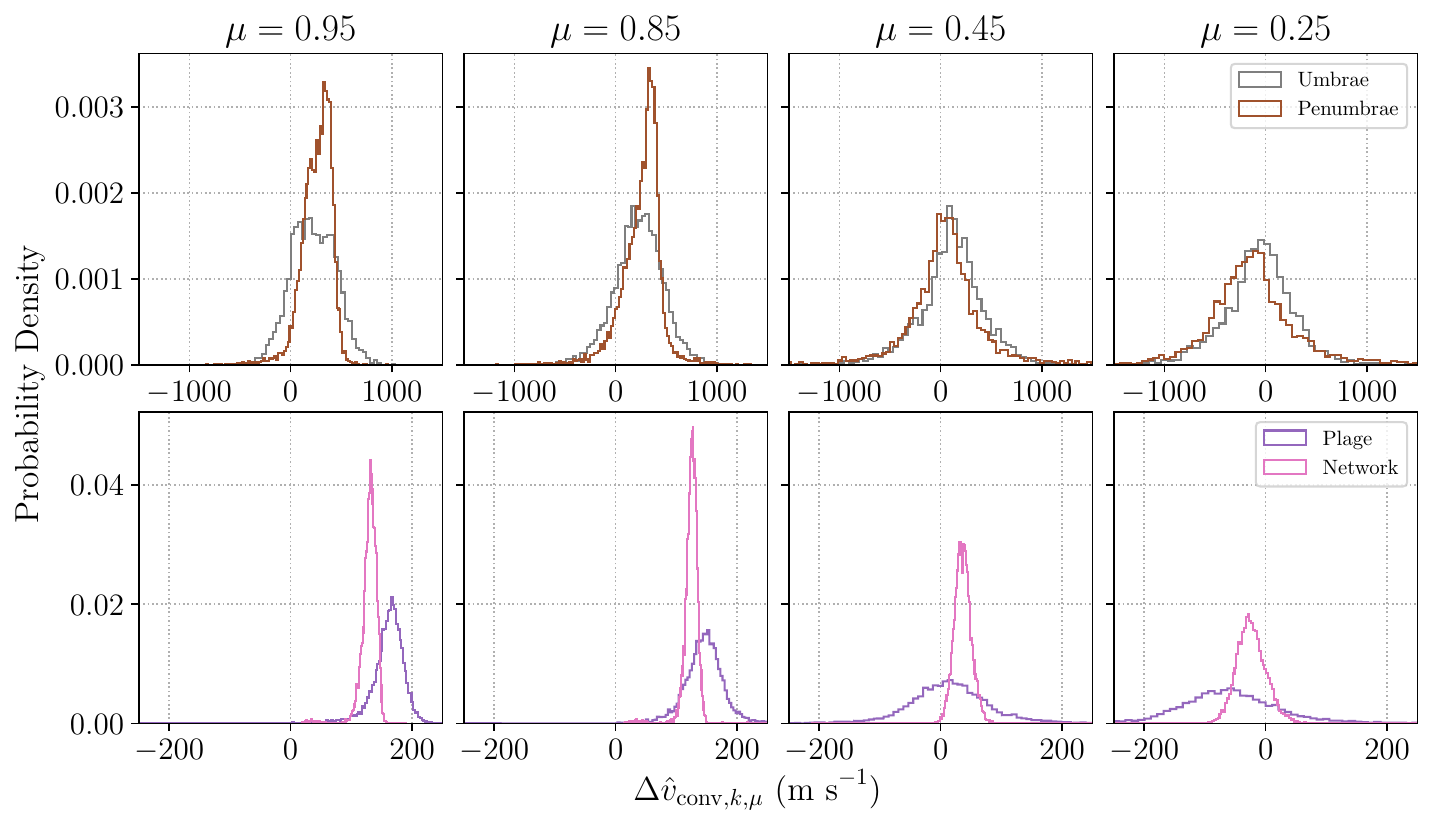}
    \caption{Distribution of $\Delta \hat{v}_{\rm conv}$ by region type (rows) for a subset of $\mu$ bins (columns). Note the different $x$- and $y$-axis scales between rows. No distribution is shown for the quiet Sun, since  by definition $\Delta \hat{v}_{\rm conv} = 0$ for this region type, as explained in \S\ref{subsec:sun_vels} 
    \textit{Top row:} The trends in the $\Delta \hat{v}_{\rm conv}$ distributions for umbrae and penumbrae generally mirror those seen in $hat{v}$ (see Figure~\ref{fig:hist}), apart from the change in distribution means.  
    \textit{Bottom row:} Likewise, the trends in the $\Delta \hat{v}_{\rm conv}$ distributions for plage and network generally follow the $\hat{v}$ distributions. However, the redshifted tail seen in the $\mu = 0.25$ panel for network in Figure~\ref{fig:hist} is notably absent in the corresponding $\Delta \hat{v}_{\rm conv}$ distribution, suggesting a continuum between magnetic network and quiet Sun convective velocities when viewed at the largest limb angles.}
    \label{fig:hist2}
    \script{fig4_5_6.py}
\end{figure*}

To quantify the center-to-limb dependence of solar RVs, we compute the velocities defined in Equations~\ref{eq:vhat}-\ref{eq:vconv} for each pixel classification described in \S\ref{sec:region_id} within several $\mu$ bins across the solar disk. These bins are linearly spaced in $\mu$, and stepped in increments of $0.1$ from $\mu = 1.0$ to $\mu = 0.1$. As described in \S\ref{sec:data}, we calculate these velocities from the SDO data sampled at a 4-hour cadence spanning 1 January 2012 to 31 December 2015, corresponding to the calendar years containing the peak of Solar Cycle 24. In \S\ref{subsec:vel_avg}, we discuss the averaged CLV trends for each region classification, which are shown in Figure~\ref{fig:clv} and tabulated in Appendix~\ref{app:clv_data_tables}. To further probe the underlying physics driving the average CLV trends, we present and analyze the full distributions of velocities seen in each region classification in \S\ref{subsec:vel_dist}. \par 

\subsection{Trends in Average Center-to-limb Velocities} \label{subsec:vel_avg}

In Figure~\ref{fig:clv}, we plot the computed velocities $\hat{v}$ and $\Delta \hat{v}_{\rm conv}$ as a function of $\mu$ for each region classification. The mean velocity for each region type is plotted as a point at the center of each $\mu$ bin, with each shaded region corresponding to the 1$\sigma$ width of the corresponding velocity distribution (see \S\ref{subsec:vel_dist} for further discussion). Because of the differences in velocity scale among the region types, we plot the velocities for umbrae and penumbrae in separate panels. The distributions of $\hat{v}$ in the top left panel of Figure~\ref{fig:clv} for quiet Sun, plage, and network all follow the general characteristic shape of the convective blueshift curve seen in \citet{Stief2019} and \citet{Cegla2018} for observed and simulated line profiles, respectively. Notably the location of the velocity minimum varies by region type. The quiet Sun has a minimum near $\mu \sim 0.7$, compared to network with its minimum at $\mu \sim 0.5$, and plage at $\mu \sim 0.45$. The 1$\sigma$ variability in velocity (as illustrated by the width of the shaded region) in plage is noticeably greater than that seen in either the quiet Sun or magnetic network in all $\mu$ bins, but especially so closer to the limb. \par 

Compared to the aforementioned regions, umbrae and penumbrae tend to produce much larger velocities, with much larger variance (see \S\ref{subsec:vel_dist}). Both umbrae and penumbrae have large, positive velocities near disk center, which taper off near the limb. In the final two bins near the limb, the mean velocity in penumbrae actually becomes slightly negative, before turning over again toward slightly positive values in the smallest $\mu$ bin. \par

The distribution of $\Delta \hat{v}_{\rm conv}$ are notably distinct from the $\hat{v}$ distributions for plage and network. Plage exhibit a strong, positive suppression of convective blueshift disk center, which gradually tapers off and inverts sign near the limb. The turnover of network velocities with limb angle is slightly less extreme with lower variance, but follows the same general trend seen in plage. \par 

Compared to plage and network, the $\Delta \hat{v}_{\rm conv}$ distributions for umbrae and penumbrae are more comparable to their corresponding $\hat{v}$ distributions. This effect is a consequence of the extreme difference in velocities between spot and quiet Sun pixels, owing to both the convective velocity that is ``missing'' in spots as well as the large velocity flows seen in penumbrae. However, as is the case for plage and network, the velocities become negative near the limb for both umbrae and penumbrae. This change in sign of the suppression of convective blueshift is discussed further in \S\ref{subsec:spot_vels}. \par

\subsection{Trends in Center-to-limb Velocity Distributions} \label{subsec:vel_dist}

Each point in Figure~\ref{fig:clv} corresponds to the mean of a whole distribution of velocities. We plot these distributions for a subset of $\mu$ bins in Figures~\ref{fig:hist} (for $\hat{v}$) and \ref{fig:hist2} (for $\Delta \hat{v}_{\rm conv}$). As hinted at by the changing $1\sigma$ width (shaded regions) seen in Figure~\ref{fig:clv}, the shapes of the velocity distributions reveal additional center-to-limb dependence. \par 

Compared to the umbrae $\hat{v}$ distribution, which remains roughly constant in width across the disk, the penumbrae distribution is relatively peaked near disk center relative to the limbward distributions. Toward the limb, the penumbrae distributions much more closely follow the umbrae distributions, apart from slight offsets in their means. \par 

The plage $\hat{v}$ distributions become very broad near the limb, whereas network velocity distributions are comparatively peaked across all $\mu$ bins. Notably, the network distribution exhibits a pronounced redshifted tail in the $\mu = 0.25$ distribution that overlaps the distribution of plage velocities for $\hat{v} \gtrsim 150\ \ms$. A similar tail in $\hat{v}$ is visible for the quiet Sun in this $\mu$ bin. \par 

The distributions of $\Delta \hat{v}_{\rm conv}$ for umbrae and penumbrae generally resemble the $\hat{v}$ distributions for these same features. As in \S\ref{sec:clv} and Figure~\ref{fig:clv}, this is a consequence of the very large difference in velocities between spot and quiet Sun pixels. \par 

As also seen in the $\hat{v}$ distributions for plage, the width of the $\Delta \hat{v}_{\rm conv}$ increases greatly with limb angle, whereas the width of the network $\Delta \hat{v}_{\rm conv}$ remains comparatively small. Notably, the redshifted tail seen in the most limbward $\Delta \hat{v}_{\rm conv}$ distribution for network appears noticeably truncated, no longer following the redshifted edge of the plage $\Delta \hat{v}_{\rm conv}$ distribution, as seen in their corresponding $\hat{v}$ distributions. As discussed in \S\ref{clv_discussion}, we interpret that the disappearance of this tail to indicate that a continuous distribution of velocities in network and quiet Sun when viewed at the most extreme limb angles. \par

\subsection{Widths of the Velocity Distributions}

\begin{figure}[!htb]
    \epsscale{1.1}
    \plotone{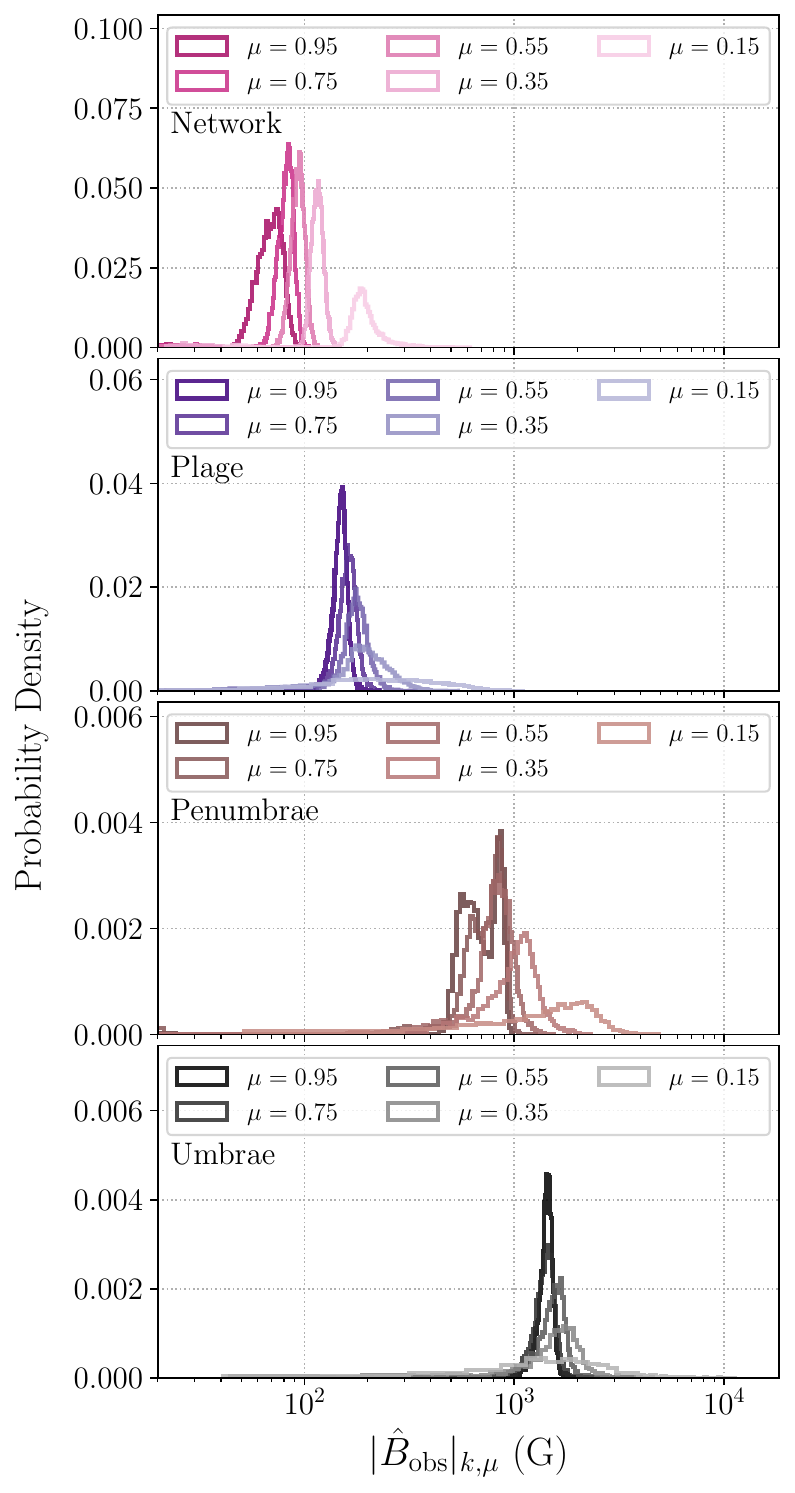}
    \caption{Distribution of unsigned magnetic flux by region classification for multiple $\mu$ bins (note the logarithmic scale on the $x$-axis). Distributions are drawn with decreasing saturation toward the limb. Generally, both the mean and the widths of the $|\hat{B}_{\rm obs}|$ distributions increase toward the limb.}
    \label{fig:b_obs_dist}
    \script{fig7_8.py}
\end{figure}

It is visually quite evident from Figures~\ref{fig:clv} and \ref{fig:hist2} that the widths of the plage and network velocity distributions vary quite differently with $\mu$. Whereas the width of the plage velocity distribution increases dramatically from the center to the limb, the network distribution is comparatively narrow at all $\mu$ bins, and the width only moderately increases from the center to the limb. Compared to plage and network, the umbrae and penumbrae velocity distributions are notably wide (hundreds of $\ms$) and also increase somewhat toward the limb. \par 

Since these regions are known to be created by magnetic fields of differing strengths and configurations \citep{Buehler2019}, these differences in widths of the velocity distributions likely reflect differences in the observed magnetic field strengths. To explore how the local solar magnetic field sculpts the plasma flows in these regions, we consider the region-averaged unsigned magnetic field flux given by:

\begin{equation}
    |\hat{B}_{\rm obs}|_{k,\mu} = \frac{\sum_{ij} |B_{\rm obs}| I_{ij} W^k_{ij}W^\mu_{ij}}{\sum_{ij} I_{ij}W^k_{ij}W^\mu_{ij}},
\end{equation}

\noindent adapting the definition of this quantity from Equation 3 of \citet{Haywood2022} for specific region classifications and $\mu$ bins, rather than the whole disk. We plot the distributions of $|\hat{B}_{\rm obs}|$ for each of network, plage, penumbrae, and umbrae at a subset of the $\mu$ bins in Figure~\ref{fig:b_obs_dist}. \par 

As expected, for a given $\mu$ bin, network are on average the least magnetic, followed by plage, then penumbrae, and lastly umbrae. For all regions, the width of the $|\hat{B}_{\rm obs}|$ distributions increases toward the limb (except for network, which has a slightly wider distribution in the $\mu =0.95$ than in the $\mu=0.75$ bin). This trend of increasing ranges of magnetic field strengths with decreasing $\mu$, would indeed suggest that the wider distributions of velocities seen toward the limb are a result of wider distributions of magnetic flux. \par

\begin{figure}[!htb]
    \epsscale{1.15}
    \plotone{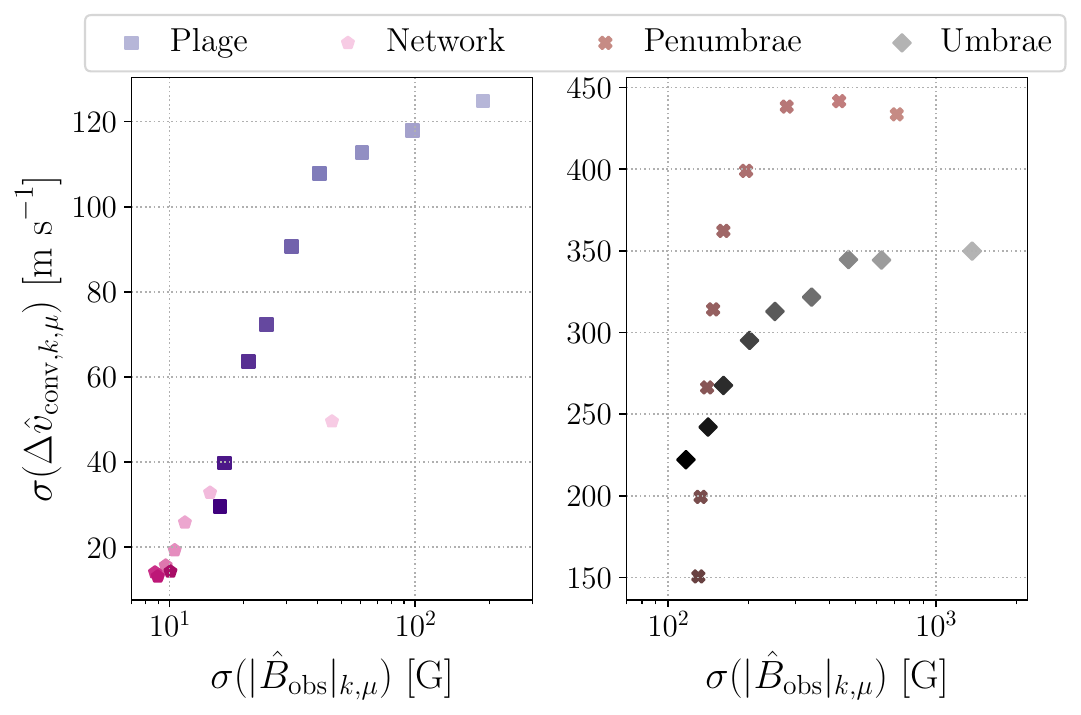}
    \caption{Wider distributions of $|\hat{B}_{\rm obs}|$ are connected with wider $\Delta\hat{v}_{\rm conv}$ distributions for all active region classifications. Note the logarithmic scale on the $x$-axis. is Each marker corresponds to a $\mu$ bin, and markers are drawn with decreasing saturation toward the limb, as in the distributions shown in Figure~\ref{fig:b_obs_dist}. }
    \label{fig:b_obs_width}
    \script{fig7_8.py}
\end{figure}

To further probe this relationship between variability in the magnetic field and variability in velocity, we plot the widths of $\Delta \hat{v}_{\rm conv}$ distributions against the widths of the $|\hat{B}_{\rm obs}|$ distributions in Figure~\ref{fig:b_obs_width}. Consistent with expectations informed by Figures~\ref{fig:clv} and \ref{fig:b_obs_dist}, wider $|\hat{B}_{\rm obs}|$ distributions are generally correlated with wider $\Delta \hat{v}_{\rm conv}$ distributions; however, this trend appears to level out at the highest magnetic flux levels, which correspond to the most limbward bins. In fact, the trend appears to reverse for the highest $|\hat{B}_{\rm obs}|$, most limbward point for the penumbrae. The penumbral distribution at low $\mu$ likely arises because the penumbral fields are to first order tangent to the surface and azimuthally symmetric around the spot center \citep{Lites1993, Tiwari2013}. Thus near the limb they should show a distribution approaching that of the cosine function, with a maximum value for small azimuth angles (measured from spot center with zero along the line-of-sight) and dropping to a fairly constant value for higher azimuth. This is very similar to the observed distribution (Figure~\ref{fig:b_obs_dist}). \par

\section{Discussion} \label{sec:discussion}

The trends seen in the center-to-limb variability (as shown in Figures~\ref{fig:clv}-\ref{fig:hist2}) are complex and distinct across the identified region types. Ultimately, these velocities arise from physical motions in the solar atmosphere, which are altered and modulated by changing magnetic fields, weighted by the local emergent intensity, and projected along our line of sight. In this section, we offer physical interpretations of the features seen in the center-to-limb velocity distributions presented in \S\ref{sec:clv}. \par

\subsection{Shape and Curvature of the CLV} \label{clv_discussion}

The characteristic curve seen in Figure~\ref{fig:clv} for the quiet Sun is well-known and attributed to variations in convective blueshift that arise from the changing physical height probed at different viewing angles. I.e., as the line of sight moves toward the limb, the absorption line contribution function shifts toward greater physical heights in the atmosphere with changing velocity structure \citep[see, e.g., the discussion of the Eddington-Barbier approximation in \S2.2.2 of][]{Rutten2003}. The exact shape of the quiet-Sun convective blueshift variation is known to vary line-to-line \citep[as shown in with spectroscopic observations in][]{Lohner-Bottcher2018b, Lohner-Bottcher2019, Stief2019} and to have sensitivity to magnetic field strength and configuration \citep[as seen in analyses of MHD simulations, e.g.,][]{Cegla2018}. \par

Compared to the mean trend seen for the quiet Sun in Figure~\ref{fig:clv}, the distribution of velocities shown in Figure~\ref{fig:hist} yields additional insight into the angular dependence of the projected convective velocities. We interpret one of these nuances, the extended, redshifted tail seen in the most limbward $\hat{v}$ distribution for the quiet Sun, as a manifestation of the ``corrugated'' nature of granules. \citet{Dravins2008} invoke an analogy that casts granules and intergranular lanes as ``hills'' and ``valleys,'' respectively \citep[see also][]{Balthasar1985}. At disk center, these hills and valleys are viewed top-down, but toward the limb, this corrugated surface is viewed more edge-on, and the hilltops can obscure pieces of this surface, increasing the apparent granule filling factor relative to the intergranular lanes. As seen in \citet{Cegla2018}, at the far limb, only the nearest edge of the granules are blueshifted, and the rest of the granule begins to point away from the line of sight. Because these blueshifted velocities at large limb angles are preferentially obscured by the tops of granules, the redshifted velocity components that sit above the intergranular lanes dominate the RV. \par 

The CLVs in $\hat{v}$ for network and plage in Figure~\ref{fig:clv} are generally reminiscent of that for the quiet Sun, but with altered curvature, vertical offsets, and minima. These differences can be attributed to alterations in the convective motion induced by the changes in local magnetic fields that create these regions. These changes in curvature and offset are seen in the \citet{Cegla2018} simulations conducted with vertical magnetic field strengths comparable to those seen in real solar network and plage. They find that the increased magnetic field strength in these regions inhibits plasma flows, pushing the minimum of the CLV curve limbward relative to the quiet Sun. A redshifted tail similar to the one seen in the $\mu = 0.25$ quiet-Sun  $\hat{v}$ distribution is present in the corresponding network velocity distribution, but is notably absent in the $\Delta \hat{v}_{\rm conv}$ distribution. The absence of this tail suggests that the horizontal components of the plasma velocity, which shape the velocity distributions near the limb, are quite similar for magnetic network and the quiet Sun; whereas the vertical velocity components, which shape the distributions at and just off disk center, are quite different for these regions.  \par 

The $\Delta \hat{v}_{\rm conv}$ CLV curves show the difference between the altered plasma motions in areas of increased magnetic field strength and the quiet-Sun magnetoconvection, i.e., the suppression of convective blueshift. It is notable that the mean value of this quantity changes sign across the solar disk for each of plage, network, umbrae, and penumbrae: i.e., the weighted sum over velocities in these regions is greater than that in quiet Sun near disk center, and less than it toward the limb. Put succinctly, the observed suppression of convective blueshift is a function of the limb angle/rotation phase of an active region. We discuss the implications of this observation in \S\ref{subsec:best_use}. \par

Compared to the quiet-Sun velocities, which are quite narrowly distributed for all $\mu$, the velocity distributions for plage and network notably increase in width from disk center to the limb. The increase in width is especially dramatic for plage, which have $\sigma(\Delta \hat{v}_{\rm conv}) \sim 30\ \ms$ at disk center and  $\sigma(\Delta \hat{v}_{\rm conv}) \sim 125\ \ms$ at the limb. As shown in Figures~\ref{fig:b_obs_dist} and \ref{fig:b_obs_width}, the increase in the width of the velocity distributions is connected to an increase in the width of the $|\hat{B}_{\rm obs}|$ distributions for both network and plage. Although the $|\hat{B}_{\rm obs}|$ distributions increase in width toward the limb for both plage and network, the increase in width is especially dramatic for plage. Since plage fields span a larger range of strengths than network fields, it is unsurprising that we see a mix of plasmas that are less and more magnetically confined and thus span a larger range of velocities than in network. \par 

\subsection{Umbral and Penumbral Velocities} \label{subsec:spot_vels}

Starspots have been long known to create large RV perturbations in disk-integrated velocities as a result of both their large flux deficit and their strong suppression of convective blueshift. In the following sections, we discuss the observed CLV trends and distributions observed for umbrae and penumbrae. \par 

\subsubsection{Umbrae}

We see velocities generally consistent with known umbral properties. Umbral magnetic fields, and thus the flows, are predominantly vertical with some spreading with height and towards the periphery. The mean velocity is largest and slightly redshifted relative to the quiet Sun at disk center, reflecting, as in plage, the dominant (but not total; see, e.g., \citealt{Lohner-Bottcher2018a}) suppression of convective blueshift. Upflows in umbral dots, which  cover only $\sim$10\% of the umbral area and are $\sim$20\% brighter than their surroundings \citep[e.g., ][]{Riethmuller+2008} likely make only a small contribution. The significant velocity dispersion is driven by waves \citep[see, e.g.,][and references therein]{Bogdan2000} and is largest near disk center and drops towards the limb, consistent with largely vertical fields channeling the waves.  The flat topped distribution is consistent with waves being driven by a regular convective ``piston." The dispersion in velocities drops off less sharply than $\mu$, however, probably due to the mild spreading of the field lines with height, which adds a tangential component towards the limb.  The tangential component likely partially accounts for the wings of the velocity distribution profile towards the limb. Also, near the limb some scattered penumbral light and mixed umbral-penumbral pixels may be affecting the velocities as well, as this would pick up stronger tangential flows seen there (see \S\ref{subsubsec:pen_disc} below). \par

\vspace{-2mm}
\subsubsection{Penumbrae} \label{subsubsec:pen_disc}

The tendency of the penumbral $\Delta \hat{v}_{\rm conv}$ to approach that of umbrae toward the limb may initially seem peculiar, owing to the naive expectation that the surface-tangent velocities observed in penumbral Evershed flows should strongly influence the shape of the penumbrae velocity distribution at larger viewing angles (i.e., low $\mu$ values). We consider three possibilities to account for the similarity between the distributions for these regions: 1) the azimuthal symmetry of penumbrae velocity flows, 2) limitations of the instrumental spatial resolution and HMI velocity-calculation algorithm, and 3) possible imperfect separation of umbrae and penumbrae by our classification scheme.  \par  

One effect that characterizes penumbrae is the presence of strong outward flows known as Evershed flows \citep{Evershed1909}. These horizontal flows achieve velocities of several hundred $\ms$ to a few $\kms$ \citep{Rimmele2006, Franz2009, Solanki2003}. Observations have shown that these flows turn downward at the outer edges of some penumbrae and upward at the inner, umbrae-bordering edges (\citealt{WestendorpPlaza1997} and \citealt{Schlichenmaier1999}; see also the review by \citealt{Solanki2003}). By area, the horizontal outward flows dominate the penumbrae and are azimuthally symmetric. As a result of this azimuthal symmetry, we expect the velocities flows normal to the solar surface to dominate the observed velocity, especially near disk center. Imperfections in this symmetry (e.g., from shielding of the near-side of the penumbra by the Wilson depression - \citealt{Wilson1774}) would lead to some fraction of the receding or approaching flows dominating the apparent RV, especially at the limb, where these motions are along the line of sight. The increase in width of the penumbrae $\Delta \hat{v}_{\rm conv}$ toward the limb could plausibly result from this effect. \par

Two additional factors introduced by the design of SDO and the calculations of the HMI observables may affect the penumbral velocities particularly strongly. First, HMI achieves an angular resolution of about 1 arcsecond, or about $\sim$700 km at disk center. This resolution is comparable to or larger than the typical widths of the high-velocity threads which create the structure of the penumbra (see \citealt{Solanki2003} for a broad overview), which are typically $\sim$0.8 arcseconds wide (at disk center) per the analysis of \citet{Tiwari2013}. Because the typical filament width falls below the resolution of HMI, the velocities that HMI measures are likely somewhat significantly blended with the surrounding penumbral structure in individual HMI pixels. In addition to the limitations introduced by the spatial sampling, it is important to recognize the limitations of the algorithm used to measure velocities from the raw HMI data. The full algorithm and caveats are described in \citet{Couvidat2016} and on the JSOC website\footnote{\url{http://jsoc.stanford.edu/relevant_papers/observables.pdf}}, but it is noted that the MDI-like algorithm is particularly prone to error in high-field strength and high-inclination regions.  \par 

\begin{figure*}[!htb]
    \epsscale{1.05}
    \plotone{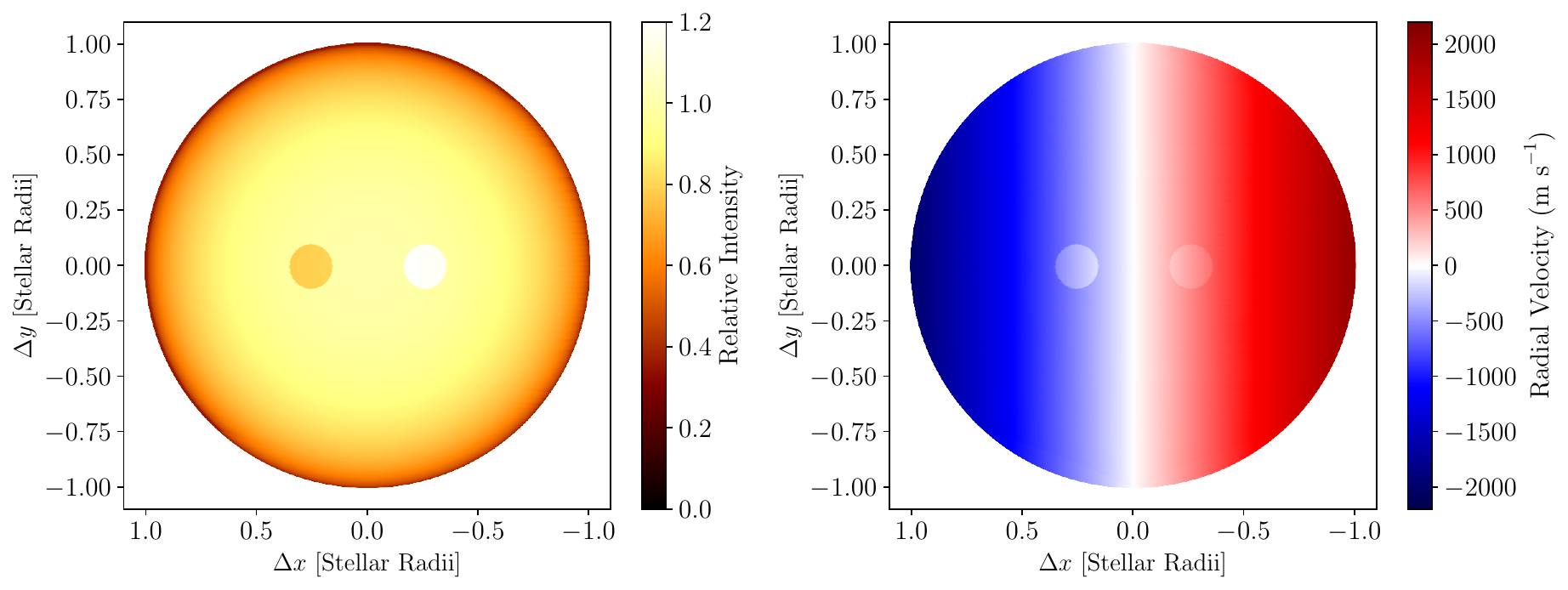}
    \caption{Visualization of the toy stellar model invoked in \S\ref{subsec:best_use}. \textit{Left:} Intensity map of the star, showing a model spot and plage. \textit{Right:} Velocity map of the model star; in addition to rotation, the spot region is downflowing, and the plage upflowing. The disk-integrated RV calculated as the intensity-weighted sum of the patch velocities differs from that calculated via fit to the disk-integrated line profile by as much as $\sim$35$\cms$, as discussed in \S\ref{subsec:best_use}.}
    \label{fig:model_star}
    \script{fig9.jl}
\end{figure*}

We also consider the possibility that our classification scheme for umbrae and penumbrae insufficiently or imperfectly discriminates between these regions, perhaps particularly near the limb. It is evident from Figure~\ref{fig:three} that there is no sharp intensity cut-off separating the brightness of penumbrae from umbrae. Rather, the low-intensity end of the HMI continuum intensity distribution is characterized by a relatively flat tail which we denote umbrae, and a gradually rising region, which we denote penumbrae. Even if the intensity threshold used to separate these regions is generally sufficient, misclassifications of even a small number of pixels could artificially drive the observed low-$\mu$ velocity distributions toward the average spot distribution. We note that an additional classification criterion, based on, for example, the presence of a large velocities as in Evershed flows, would not function near disk center (where these velocities are perpendicular to the line of sight) and would be fraught near the limb (where the aforementioned limitations of the velocity-calculation algorithm are most significant). \par 

\vspace{-1mm}
\subsection{Application of CLV Trends} \label{subsec:best_use}

We emphasize that the velocities and trends presented in this work cannot be trivially scaled (e.g., by apparent intensity and filling factor of a feature at a given phase) to yield the expected RV deviation measured observationally. This can work in some cases, but is not typically true. To demonstrate this effect, we numerically model a toy stellar disk from which we calculate disk-integrated velocities. We assume that each patch of the stellar disk produces a Gaussian line profile with continuum intensity given by a quadratic limb darkening law as in Equation~\ref{eq:limb_dark} (with $c_1$ = 0.4 and $c_2$ = 0.26), fractional depth 0.5, thermal broadening (for Fe I, as in the Fe I 6173 \AA\ line observed by HMI) of $v_{\rm therm} = $1.25 $\kms$, and microturbulence $v_{\rm mic} = 1 \kms$. The combined velocity width of the \textit{local} line profiles, given by a simple quadrature sum, is then $v_{\rm local} \sim 1.60 \kms$. \par 

To produce a disk-integrated line profile we assume that each local line profile is shifted by the local, line-of-sight-projected rotational velocity of the star. Following the example of the Sun, we set $v\sin i = 2 \kms$, with $i = 90$ degrees. For the sake of simplicity, we assume solid-body rotation. The integrated line profile is then given by the $\mu$- and intensity-weighted sum over the local line profiles, which are convolved with a Gaussian macroturbulence kernel with width $v_{\rm mac} = 3 \kms$ (as in Equation 17.10 of \citealt{Gray2008}). Because the rotational broadening is not modeled as an additional Gaussian broadening kernel, the final width of the disk-integrated line is somewhat narrower ($v_{\rm tot} \sim 3.54 \kms$) than the width predicted by a simple quadrature sum over all broadening sources ($\sim$3.9 $\kms$).

We calculate a disk-integrated radial velocity for this model star in three ways: 1) as the intensity-weighted sum over the individual patch velocities, 2) by fitting a Gaussian to the disk-integrated line profile, and 3) by fitting a parabola to the bottom 5\% of the disk-integrated line profile. For this first stellar model, which lacks any perturbations from spots, etc., each of the velocity-measurement methods yield an RV of $\sim$0 $\ms$. To model the effects of stellar activity, we now introduce simple model starspot and plage, as shown in Figure~\ref{fig:model_star}. The starspot is dark ($I_{\rm spot} = 0.8 I_{\rm quiet}$), downflowing ($v_{\rm spot} = +200$ $\ms$), and centered at stellar latitude $0$ degrees and longitude $+15$ degrees from disk center. The plage is bright ($I_{\rm plage} = 1.2 I_{\rm quiet}$), upflowing ($v_{\rm plage} = -100$ $\ms$), and centered at $0$ degrees latitude and $-15$ degrees longitude from disk center. Both regions occupy 0.5\% of the surface area of the visible hemisphere of the star. With this configuration, the weighted-sum velocity is $\sim$2.81 $\ms$; the velocity measured from the Gaussian fit is $\sim$2.94 $\ms$; and the velocity measured via quadratic fit to the line core is $\sim$3.16 $\ms$. These velocities differ by $\sim$13$\cms$ at best, and $\sim$35$\cms$ at worst. \par 

The mismatch between the RVs obtained via these methods is driven by a few factors. First, the local, disk-resolved line profiles are not as broad as disk-integrated stellar line profiles (due to rotation at the very least). Second, ``imbalances'' in flux and velocity created by stellar-surface inhomogeneities create asymmetries in the final disk-integrated line profile, even if the local profiles are assumed to be symmetric (as in this simplistic toy model). Consequently, different methods of fitting the resulting line profile (e.g., Gaussian fit to the whole line vs.\@ a Gaussian or quadratic fit to only the line core) will produce different results. This insight suggests that optimal use of the CLV trends presented in this work may require using them to model an ``average" stellar line (or a spectrum) and derive the ``observed" velocity by fitting to this ``average'' line. \par

\vspace{-1mm}
\subsection{Implications for Extremely Precise Radial Velocities and Future Directions} \label{subsec:future_direct}

Direct observations and analyses of the complex spatial behaviors exhibited by the various region types on the Sun are not available for other stars, owing to our current inability to resolve (at high spatial resolution) the surfaces of slowly rotating, relatively inactive stars that are commonly selected for RV surveys. Precisely for this reason, observing and understanding these trends as they manifest on the Sun, is a key step in building physically motivated models of intrinsic stellar variability that are necessary to identify the signals of low-mass planets in RV observations, and to determine their masses accurately \citep[see the Executive Summary and \S A2 of][]{Crass2021}. Moreover, greater physical insight into these activity signals is necessary to inform the application of advanced statistical models (e.g., Gaussian Processes) such that real planetary signals are not inadvertently absorbed into the stellar noise components. Future iterations of stellar activity simulations should consider the limb-angle and region-dependent nature of the suppression of convective blueshift observed in this work. Much like spatially-resolved spectroscopic observations of the Sun have been used to construct (semi-)empirical models of stellar variability at the spectral level \citep[as in][]{Palumbo2022, Zhao2023}, other future works could use SDO data as building blocks of detailed, empirical simulations of stellar surfaces. However, as shown in \S\ref{subsec:best_use} and discussed in \S\ref{subsubsec:spectrographs}, caution should be taken to account for differences in how solar vs.\@ stellar, and disk-resolved vs.\@ disk-integrated velocities are modeled and measured. \par 

We emphasize that the analyses of the CLV trends presented in this work are by no means complete. Notably, we do not explicitly quantify the center-to-limb behavior of other stellar phenomena that are suspected impact the measurement of precise RVs \citep[see \S 6 of][]{Haywood2022}, including meridional circulations, moat flows \citep{Solanki2003}, active region inflows \citep{Gizon2001, Gizon2010}, ``bright grains" in penumbral Evershed flows \citep{Rimmele2006}, and flares \citep{Reiners2009, Saar2018}. Ostensibly, these sources of variability are present in the SDO data analyzed herein, but their signatures have not been isolated and analyzed separately from the pixel classifications used in this work. Future works could attempt to isolate these phenomenon, characterize their apparent velocities as a function of limb angle, and assess their contribution to the total Sun-as-a-star RV. \par 

\subsubsection{Need for Line-by-Line Characterizations of Convective Blueshift} \label{subsubsec:spectrographs}

It is vital to recognize that the center-to-limb trends studied in this work are derived from observations of a single absorption line, whereas velocities measured in RV observations of other stars are produced from measurements of many thousands of lines. Moreover, the method by which velocities are measured from HMI observations (see the discussion in \S\ref{subsubsec:pen_disc} and the description of the MDI-like algorithm in \citealt{Couvidat2016}) varies fundamentally from the various velocity-measurement algorithms employed by RV spectrograph data-reduction pipelines (see, e.g., \citealt{Pepe2002} for a description of the CCF method and \citealt{Petersburg2020} for an example of a forward-modeling method). \par

Previous works (including \citealt{Haywood2016}, \citealt{Milbourne2019}, \citealt{Ervin2022}, and \citealt{Haywood2022}) attempted to account for these different RV measurement prescriptions by determining linear best-fit coefficients which could be used to transform HMI-derived velocities into model RV variations like those that would be measured by a ground-based spectrograph (see \S2.10 of \citealt{Ervin2022}). Although this modeling method led to important insights, such as the identification of the disk-averaged unsigned magnetic flux as a powerful proxy for RV variations by \citet{Haywood2022}, studies of (the suppression of) convective blueshift for individual lines will be important for fully understanding stellar variability in the context of EPRVs. This is because different absorption lines are known to trace different components of the full 3D convective velocity field, and consequently exhibit different center-to-limb trends in their shapes \citep{Lohner-Bottcher2018b, Lohner-Bottcher2019} and variability \citep{AlMoulla2022}. Assuming line-by-line methods of RV extraction (e.g., \citealt{Dumusque2018}) are better equipped to cope with this differential line jitter, additional studies of variability at the level of individual lines will be needed to fully disentangle variability in convective blueshift from Doppler shifts arising from bulk stellar motion. \par 

\vspace{-2mm}
\section{Conclusion}

In this work, we use observables from SDO/HMI and SDO/AIA to explore the velocity center-to-limb variability (CLV) for a subset of solar surface features, including quiet Sun, umbrae, penumbrae, network, and plage. We:

\begin{itemize}
    \setlength\itemsep{0.02em}
    \item find that the SDO data provide sufficient information to robustly classify solar features at $\sim$arcsecond resolution, including separate identifications of sunspot umbrae and penumbrae (Figures \ref{fig:one}, \ref{fig:two}, and \ref{fig:three}),
    \item recover the CLV signature of the quiet Sun, magnetic network, plage, umbrae, and penumbrae (Figure \ref{fig:clv}), 
    \item show that the shape of the velocity distributions for these regions are strongly dependent on limb angle (Figures~\ref{fig:hist} and \ref{fig:hist2}), 
    \item verify that our methodology reproduces the known center-to-limb dependence of the convective blueshift observed in the quiet Sun (\S\ref{clv_discussion}), 
    \item show that the suppression of convective blueshift observed in magnetic network, plage, umbrae, and penumbrae vary in strength and sign as a function of limb angle (right-hand panels of Figure~\ref{fig:clv} and \S\ref{clv_discussion}), 
    \item emphasize that the limb-angle dependencies shown in this work cannot be trivially scaled to yield the RV deviation that would be measured from disk-integrated absorption lines (\S\ref{subsec:best_use}),
    \item and recommend that future stellar activity simulations and statistical models consider these region-dependent, limb-angle trends and variances (\S\ref{subsec:future_direct}).
\end{itemize}

\section{Acknowledgements}
\noindent We thank the anonymous referee whose review has improved the content and clarity of this work. M.L.P.\@ warmly thanks H.M.\@ Cegla, A.\@ Mortier, and N.C.\@ Santos for conversations and insights that have improved this work. SDO data products used in this work are provided courtesy of NASA/SDO and the AIA, EVE, and HMI science teams. M.L.P.\@ acknowledges the support of the Penn State Academic Computing Fellowship. The Center for Exoplanets and Habitable Worlds is supported by the Pennsylvania State University and the Eberly College of Science. Computations for this research were performed on the Pennsylvania State University’s Institute for Computational and Data Sciences’ Roar supercomputer. S.H.S.\@ gratefully acknowledges support by NASA EPRV grant 80NSSC21K1037, XRP grant 80NSSC21K0607, and Heliophysics LWS grant NNX16AB79G. R.D.H.\@ is funded by the UK Science and Technology Facilities Council (STFC)'s Ernest Rutherford Fellowship (grant number ST/V004735/1). This work was performed in part under contract with the California Institute of Technology (Caltech)/Jet Propulsion Laboratory (JPL) funded by NASA through the Sagan Fellowship Program (Grant number 1581003) executed by the NASA Exoplanet Science Institute (R.D.H.). This research has made use of NASA's Astrophysics Data System Bibliographic Services.

\facilities{SDO}

\software{Astropy \citep{AstropyCollaboration2013,AstropyCollaboration2018,AstropyCollaboration2022}, GNU Parallel \citep{tange_2021_5523272}, Matplotlib (\citealt{Hunter2007}), NumPy \citep{harris2020array}, pandas \citep{mckinney2010data}, scikit-image \citep{van2014scikit}, SciPy \citep{2020SciPy-NMeth}, SunPy \citep{SunPyCommunity2020}}

\clearpage
\newpage
\appendix
\section{Dopplergram Corrections and Systematics} \label{app:vel_components}

\begin{figure*}[!htb]
    \epsscale{1}
    \plotone{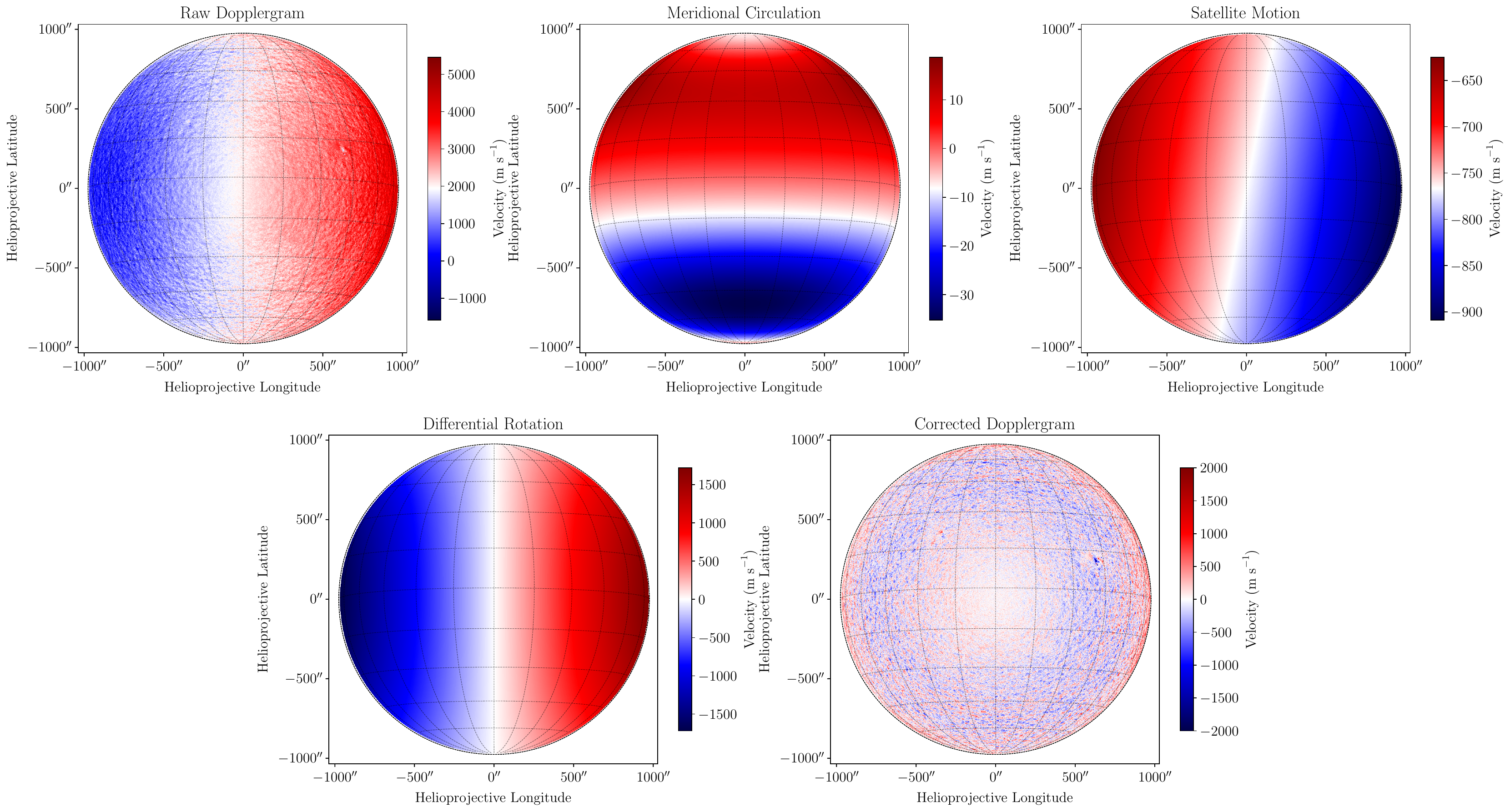}
    \caption{Example velocity fields for the Dopplergram correction terms in Equation~\ref{eq:v_corr}. Gravitational redshift and the convective blueshift terms are not shown because they are constant across all pixel values, as described in \S\ref{subsec:processing}. Note that the color scale used for velocity differs in each panel. The velocity scale for the corrected Dopplergram (bottom right panel) matches that of Figure~\ref{fig:one}, Panel (c).}
    \label{fig:vel_components}
    \script{fig10.py}
\end{figure*}

To compute Dopplergrams corrected for satellite motion, differential rotation, meridional circulation, gravitational redshift, and a convective blueshift offset, we adapt the methods from \citet{Haywood2016}, \citet{Kashyap2021}, and \citet{Ervin2022}, as explained in \S\ref{subsec:processing}. Example visualizations of these various terms are shown in Figure~\ref{fig:vel_components}, except for gravitational redshift and the convective blueshift offset, the values of which are constant for all pixels. It is important to note that these velocity corrections alone cannot account for the various systematics known to exist in HMI Dopplergrams, which are due to uncertainty in the precise orbital motion of the spacecraft, fringing from the instrument front window (which acts as a weak Fabry-P\'{e}rot interferometer), other thermal perturbations, etc.\@ (as described in detail in \citealt{Couvidat2016}). \par 

Historically, works which have used SDO/HMI data to investigate Sun-as-a-star velocities have relied on a few techniques to overcome these systematics. These techniques are described in the greatest detail in \S2.1.2 of \citet{Haywood2022}. In summary, we account for the lack of a long-timescale absolute velocity calibration by measuring the suppression of convective blueshift relative to the (region-averaged) quiet Sun velocity, as shown in Equation~\ref{eq:vconv}. We account for the uncertainties resulting from orbital velocity and daily temperature variations by averaging over multiple (in our case, six) daily observations. As noted in \S\ref{subsubsec:spectrographs}, previous works (including \citealt{Haywood2016}, \citealt{Milbourne2019}, \citealt{Ervin2022}, and \citealt{Haywood2022}) that have examined time-series Sun-as-a-star velocities from SDO/HMI data have also accounted for other systematic differences between SDO and ground-based RV spectrographs by constructing their final $\Delta$RV as the linear combination of the HMI-derived flux-effect and suppression-of-convective-blueshift velocities (see Equation 16 of Ervin et al. 2022), for which coefficients where determined via least-squares optimization using ground-based RV measurements from HARPS, HARPS-N, and/or NEID (depending on the work). We do not perform such an analysis herein, since this work is not concerned with identifying proxies for deviations in RV time series, but rather the time-average center-to-limb variations in the suppression of convective blueshift. \par 

\section{CLV Data Tables} \label{app:clv_data_tables}

Tables~\ref{tab:v_hat} and \ref{tab:v_conv} report the mean and standard deviation of the $\hat{v}$ and $\Delta \hat{v}_{\rm conv}$ distributions, respectively, shown in various forms in Figures~\ref{fig:clv}, \ref{fig:hist}, and \ref{fig:hist2}. Although \citet{Kashyap2021} fit a fifth-order polynomial to the convective blueshift trend (i.e., the quiet Sun curve in Figure~\ref{fig:clv}), the choice of functional form used to fit  these center-to-limb trends in general would be arbitrary; consequently, we refrain from fitting these data and instead directly provide the measured velocities. \par 

\vspace{10mm}
\centerwidetable
\begin{deluxetable}{c|cc|cc|cc|cc|cc}[!htb]
\tabletypesize{\footnotesize}
\tablecaption{Data values for the center-to-limb trends in $\hat{v}$ plotted at left in Figure~\ref{fig:clv}. The reported $\mu$ corresponds to the center of a given bin (e.g., $\mu$ of 0.95 corresponds to $0.9 < \mu \leq 1.0$, etc.). The reported velocities are averaged over the four-year period considered, and so are reported to$\cms$ precision. However, as discussed in \S2.1.2 of \citet{Haywood2022}, the single-measurement precision of HMI-derived RVs is (conservatively) uncertain at the level of $\sim$0.1 $\ms$. \label{tab:v_hat}}
\tablenum{1}

\tablehead{\colhead{} & \multicolumn{2}{c}{Quiet Sun} & \multicolumn{2}{c}{Network} & \multicolumn{2}{c}{Plage} & \multicolumn{2}{c}{Umbrae} & \multicolumn{2}{c}{Penumbrae} \\
$\mu$ & Mean $\hat{v}$ & $\sigma(\hat{v})$ & Mean $\hat{v}$ & $\sigma(\hat{v})$ & Mean $\hat{v}$ & $\sigma(\hat{v})$ & Mean $\hat{v}$ & $\sigma(\hat{v})$ & Mean $\hat{v}$ & $\sigma(\hat{v})$\\
& ($\ms$) & ($\ms$) & ($\ms$) & ($\ms$) & ($\ms$) & ($\ms$) & ($\ms$) & ($\ms$) & ($\ms$) & ($\ms$)}

\startdata
0.15 & 213.95 & 32.34 & 160.19 & 65.92 & 131.92 & 128.30 & 52.50 & 334.69 & -2.07 & 425.47 \\
0.25 & 95.06 & 21.34 & 72.33 & 43.04 & 39.14 & 120.16 & 29.34 & 335.16 & -33.70 & 437.54 \\
0.35 & 20.31 & 14.71 & 29.98 & 31.35 & 3.64 & 113.06 & 65.38 & 339.08 & -11.17 & 436.00 \\
0.45 & -21.66 & 11.49 & 15.36 & 22.17 & -2.44 & 107.81 & 89.05 & 318.73 & 25.67 & 398.03 \\
0.55 & -42.22 & 9.30 & 17.60 & 16.59 & 7.27 & 90.37 & 116.74 & 311.32 & 72.82 & 361.85 \\
0.65 & -49.15 & 8.43 & 33.19 & 13.22 & 31.60 & 71.25 & 140.23 & 295.87 & 124.97 & 314.52 \\
0.75 & -44.72 & 8.93 & 60.46 & 15.74 & 66.71 & 63.05 & 170.57 & 270.41 & 179.96 & 267.38 \\
0.85 & -25.22 & 12.00 & 99.14 & 18.62 & 120.09 & 41.95 & 203.10 & 246.60 & 230.54 & 201.69 \\
0.95 & 8.57 & 11.96 & 138.08 & 18.35 & 169.25 & 31.83 & 233.56 & 223.26 & 265.32 & 151.92 \\
\enddata

\end{deluxetable} 
\centerwidetable
\begin{deluxetable}{c|cc|cc|cc|cc}[!htb]
\tabletypesize{\footnotesize}
\tablecaption{Same as Table~\ref{tab:v_hat}, except for $\Delta \hat{v}_{\rm conv}$. By definition $\Delta \hat{v}_{\rm conv}$ is 0 for the quiet Sun, as explained in \S\ref{subsec:sun_vels}. \label{tab:v_conv}}
\tablenum{2}

\tablehead{\colhead{} & \multicolumn{2}{c}{Network} & \multicolumn{2}{c}{Plage} & \multicolumn{2}{c}{Umbrae} & \multicolumn{2}{c}{Penumbrae} \\
$\mu$ & Mean $\Delta \hat{v}_{\rm conv}$ & $\sigma(\Delta \hat{v}_{\rm conv})$ & Mean $\Delta \hat{v}_{\rm conv}$ & $\sigma(\Delta \hat{v}_{\rm conv})$ & Mean $\Delta \hat{v}_{\rm conv}$ & $\sigma(\Delta \hat{v}_{\rm conv})$ & Mean $\Delta \hat{v}_{\rm conv}$ & $\sigma(\Delta \hat{v}_{\rm conv})$ \\
& ($\ms$) & ($\ms$) & ($\ms$) & ($\ms$) & ($\ms$) & ($\ms$) & ($\ms$) & ($\ms$)}

\startdata
0.15 & -53.76 & 49.57 & -82.47 & 124.81 & -161.60 & 349.85 & -216.29 & 433.61 \\
0.25 & -22.74 & 32.83 & -56.14 & 117.88 & -65.30 & 344.38 & -128.93 & 441.68 \\
0.35 & 9.67 & 25.82 & -16.70 & 112.68 & 44.70 & 344.70 & -31.76 & 438.36 \\
0.45 & 37.01 & 19.30 & 19.24 & 107.77 & 110.51 & 321.65 & 47.32 & 399.01 \\
0.55 & 59.81 & 15.70 & 49.58 & 90.66 & 159.09 & 312.90 & 115.11 & 362.23 \\
0.65 & 82.34 & 13.51 & 80.78 & 72.30 & 189.38 & 295.19 & 174.19 & 314.19 \\
0.75 & 105.17 & 14.09 & 111.50 & 63.67 & 215.56 & 267.60 & 224.75 & 266.36 \\
0.85 & 124.36 & 13.14 & 145.31 & 39.76 & 228.84 & 242.13 & 255.76 & 199.39 \\
0.95 & 129.50 & 14.31 & 160.66 & 29.53 & 225.24 & 222.10 & 256.74 & 150.64 \\
\enddata

\end{deluxetable} 
\clearpage
\newpage
\bibliography{bib, misc}

\begin{thebibliography}{}
\expandafter\ifx\csname natexlab\endcsname\relax\def\natexlab#1{#1}\fi
\providecommand{\url}[1]{\href{#1}{#1}}
\providecommand{\dodoi}[1]{doi:~\href{http://doi.org/#1}{\nolinkurl{#1}}}
\providecommand{\doeprint}[1]{\href{http://ascl.net/#1}{\nolinkurl{http://ascl.net/#1}}}
\providecommand{\doarXiv}[1]{\href{https://arxiv.org/abs/#1}{\nolinkurl{https://arxiv.org/abs/#1}}}

\bibitem[{{Al Moulla} {et~al.}(2022){Al Moulla}, {Dumusque}, {Cretignier},
  {Zhao}, \& {Valenti}}]{AlMoulla2022}
{Al Moulla}, K., {Dumusque}, X., {Cretignier}, M., {Zhao}, Y., \& {Valenti},
  J.~A. 2022, \aap, 664, A34, \dodoi{10.1051/0004-6361/202243276}

\bibitem[{{Al Moulla} {et~al.}(2023){Al Moulla}, {Dumusque}, {Figueira}, {Lo
  Curto}, {Santos}, \& {Wildi}}]{AlMoulla2023}
{Al Moulla}, K., {Dumusque}, X., {Figueira}, P., {et~al.} 2023, \aap, 669, A39,
  \dodoi{10.1051/0004-6361/202244663}

\bibitem[{{Astropy Collaboration} {et~al.}(2013){Astropy Collaboration},
  {Robitaille}, {Tollerud}, {Greenfield}, {Droettboom}, {Bray}, {Aldcroft},
  {Davis}, {Ginsburg}, {Price-Whelan}, {Kerzendorf}, {Conley}, {Crighton},
  {Barbary}, {Muna}, {Ferguson}, {Grollier}, {Parikh}, {Nair}, {Unther},
  {Deil}, {Woillez}, {Conseil}, {Kramer}, {Turner}, {Singer}, {Fox}, {Weaver},
  {Zabalza}, {Edwards}, {Azalee Bostroem}, {Burke}, {Casey}, {Crawford},
  {Dencheva}, {Ely}, {Jenness}, {Labrie}, {Lim}, {Pierfederici}, {Pontzen},
  {Ptak}, {Refsdal}, {Servillat}, \& {Streicher}}]{AstropyCollaboration2013}
{Astropy Collaboration}, {Robitaille}, T.~P., {Tollerud}, E.~J., {et~al.} 2013,
  \aap, 558, A33, \dodoi{10.1051/0004-6361/201322068}

\bibitem[{{Astropy Collaboration} {et~al.}(2018){Astropy Collaboration},
  {Price-Whelan}, {Sip{\H{o}}cz}, {G{\"u}nther}, {Lim}, {Crawford}, {Conseil},
  {Shupe}, {Craig}, {Dencheva}, {Ginsburg}, {VanderPlas}, {Bradley},
  {P{\'e}rez-Su{\'a}rez}, {de Val-Borro}, {Aldcroft}, {Cruz}, {Robitaille},
  {Tollerud}, {Ardelean}, {Babej}, {Bach}, {Bachetti}, {Bakanov}, {Bamford},
  {Barentsen}, {Barmby}, {Baumbach}, {Berry}, {Biscani}, {Boquien}, {Bostroem},
  {Bouma}, {Brammer}, {Bray}, {Breytenbach}, {Buddelmeijer}, {Burke},
  {Calderone}, {Cano Rodr{\'\i}guez}, {Cara}, {Cardoso}, {Cheedella}, {Copin},
  {Corrales}, {Crichton}, {D'Avella}, {Deil}, {Depagne}, {Dietrich}, {Donath},
  {Droettboom}, {Earl}, {Erben}, {Fabbro}, {Ferreira}, {Finethy}, {Fox},
  {Garrison}, {Gibbons}, {Goldstein}, {Gommers}, {Greco}, {Greenfield},
  {Groener}, {Grollier}, {Hagen}, {Hirst}, {Homeier}, {Horton}, {Hosseinzadeh},
  {Hu}, {Hunkeler}, {Ivezi{\'c}}, {Jain}, {Jenness}, {Kanarek}, {Kendrew},
  {Kern}, {Kerzendorf}, {Khvalko}, {King}, {Kirkby}, {Kulkarni}, {Kumar},
  {Lee}, {Lenz}, {Littlefair}, {Ma}, {Macleod}, {Mastropietro}, {McCully},
  {Montagnac}, {Morris}, {Mueller}, {Mumford}, {Muna}, {Murphy}, {Nelson},
  {Nguyen}, {Ninan}, {N{\"o}the}, {Ogaz}, {Oh}, {Parejko}, {Parley}, {Pascual},
  {Patil}, {Patil}, {Plunkett}, {Prochaska}, {Rastogi}, {Reddy Janga},
  {Sabater}, {Sakurikar}, {Seifert}, {Sherbert}, {Sherwood-Taylor}, {Shih},
  {Sick}, {Silbiger}, {Singanamalla}, {Singer}, {Sladen}, {Sooley},
  {Sornarajah}, {Streicher}, {Teuben}, {Thomas}, {Tremblay}, {Turner},
  {Terr{\'o}n}, {van Kerkwijk}, {de la Vega}, {Watkins}, {Weaver}, {Whitmore},
  {Woillez}, {Zabalza}, \& {Astropy Contributors}}]{AstropyCollaboration2018}
{Astropy Collaboration}, {Price-Whelan}, A.~M., {Sip{\H{o}}cz}, B.~M., {et~al.}
  2018, \aj, 156, 123, \dodoi{10.3847/1538-3881/aabc4f}

\bibitem[{{Astropy Collaboration} {et~al.}(2022){Astropy Collaboration},
  {Price-Whelan}, {Lim}, {Earl}, {Starkman}, {Bradley}, {Shupe}, {Patil},
  {Corrales}, {Brasseur}, {N{\"o}the}, {Donath}, {Tollerud}, {Morris},
  {Ginsburg}, {Vaher}, {Weaver}, {Tocknell}, {Jamieson}, {van Kerkwijk},
  {Robitaille}, {Merry}, {Bachetti}, {G{\"u}nther}, {Aldcroft},
  {Alvarado-Montes}, {Archibald}, {B{\'o}di}, {Bapat}, {Barentsen},
  {Baz{\'a}n}, {Biswas}, {Boquien}, {Burke}, {Cara}, {Cara}, {Conroy},
  {Conseil}, {Craig}, {Cross}, {Cruz}, {D'Eugenio}, {Dencheva}, {Devillepoix},
  {Dietrich}, {Eigenbrot}, {Erben}, {Ferreira}, {Foreman-Mackey}, {Fox},
  {Freij}, {Garg}, {Geda}, {Glattly}, {Gondhalekar}, {Gordon}, {Grant},
  {Greenfield}, {Groener}, {Guest}, {Gurovich}, {Handberg}, {Hart},
  {Hatfield-Dodds}, {Homeier}, {Hosseinzadeh}, {Jenness}, {Jones}, {Joseph},
  {Kalmbach}, {Karamehmetoglu}, {Ka{\l}uszy{\'n}ski}, {Kelley}, {Kern},
  {Kerzendorf}, {Koch}, {Kulumani}, {Lee}, {Ly}, {Ma}, {MacBride}, {Maljaars},
  {Muna}, {Murphy}, {Norman}, {O'Steen}, {Oman}, {Pacifici}, {Pascual},
  {Pascual-Granado}, {Patil}, {Perren}, {Pickering}, {Rastogi}, {Roulston},
  {Ryan}, {Rykoff}, {Sabater}, {Sakurikar}, {Salgado}, {Sanghi}, {Saunders},
  {Savchenko}, {Schwardt}, {Seifert-Eckert}, {Shih}, {Jain}, {Shukla}, {Sick},
  {Simpson}, {Singanamalla}, {Singer}, {Singhal}, {Sinha}, {Sip{\H{o}}cz},
  {Spitler}, {Stansby}, {Streicher}, {{\v{S}}umak}, {Swinbank}, {Taranu},
  {Tewary}, {Tremblay}, {de Val-Borro}, {Van Kooten}, {Vasovi{\'c}}, {Verma},
  {de Miranda Cardoso}, {Williams}, {Wilson}, {Winkel}, {Wood-Vasey}, {Xue},
  {Yoachim}, {Zhang}, {Zonca}, \& {Astropy Project
  Contributors}}]{AstropyCollaboration2022}
{Astropy Collaboration}, {Price-Whelan}, A.~M., {Lim}, P.~L., {et~al.} 2022,
  \apj, 935, 167, \dodoi{10.3847/1538-4357/ac7c74}

\bibitem[{{Balthasar}(1985)}]{Balthasar1985}
{Balthasar}, H. 1985, \solphys, 99, 31, \dodoi{10.1007/BF00157296}

\bibitem[{{Bogdan}(2000)}]{Bogdan2000}
{Bogdan}, T.~J. 2000, \solphys, 192, 373, \dodoi{10.1023/A:1005225214520}

\bibitem[{{Buehler} {et~al.}(2019){Buehler}, {Lagg}, {van Noort}, \&
  {Solanki}}]{Buehler2019}
{Buehler}, D., {Lagg}, A., {van Noort}, M., \& {Solanki}, S.~K. 2019, \aap,
  630, A86, \dodoi{10.1051/0004-6361/201833585}

\bibitem[{{Cavallini} {et~al.}(1985){Cavallini}, {Ceppatelli}, \&
  {Righini}}]{Cavallini1985}
{Cavallini}, F., {Ceppatelli}, G., \& {Righini}, A. 1985, \aap, 143, 116

\bibitem[{{Cegla} {et~al.}(2019){Cegla}, {Watson}, {Shelyag}, {Mathioudakis},
  \& {Moutari}}]{Cegla2019}
{Cegla}, H.~M., {Watson}, C.~A., {Shelyag}, S., {Mathioudakis}, M., \&
  {Moutari}, S. 2019, \apj, 879, 55, \dodoi{10.3847/1538-4357/ab16d3}

\bibitem[{{Cegla} {et~al.}(2018){Cegla}, {Watson}, {Shelyag}, {Chaplin},
  {Davies}, {Mathioudakis}, {Palumbo}, {Saar}, \& {Haywood}}]{Cegla2018}
{Cegla}, H.~M., {Watson}, C.~A., {Shelyag}, S., {et~al.} 2018, \apj, 866, 55,
  \dodoi{10.3847/1538-4357/aaddfc}

\bibitem[{{Cosentino} {et~al.}(2014){Cosentino}, {Lovis}, {Pepe}, {Collier
  Cameron}, {Latham}, {Molinari}, {Udry}, {Bezawada}, {Buchschacher},
  {Figueira}, {Fleury}, {Ghedina}, {Glenday}, {Gonzalez}, {Guerra}, {Henry},
  {Hughes}, {Maire}, {Motalebi}, \& {Phillips}}]{Cosentino2014}
{Cosentino}, R., {Lovis}, C., {Pepe}, F., {et~al.} 2014, in Society of
  Photo-Optical Instrumentation Engineers (SPIE) Conference Series, Vol. 9147,
  Ground-based and Airborne Instrumentation for Astronomy V, ed. S.~K.
  {Ramsay}, I.~S. {McLean}, \& H.~{Takami}, 91478C, \dodoi{10.1117/12.2055813}

\bibitem[{{Couvidat} {et~al.}(2016){Couvidat}, {Schou}, {Hoeksema}, {Bogart},
  {Bush}, {Duvall}, {Liu}, {Norton}, \& {Scherrer}}]{Couvidat2016}
{Couvidat}, S., {Schou}, J., {Hoeksema}, J.~T., {et~al.} 2016, \solphys, 291,
  1887, \dodoi{10.1007/s11207-016-0957-3}

\bibitem[{{Crass} {et~al.}(2021){Crass}, {Gaudi}, {Leifer}, {Beichman},
  {Bender}, {Blackwood}, {Burt}, {Callas}, {Cegla}, {Diddams}, {Dumusque},
  {Eastman}, {Ford}, {Fulton}, {Gibson}, {Halverson}, {Haywood}, {Hearty},
  {Howard}, {Latham}, {L{\"o}hner-B{\"o}ttcher}, {Mamajek}, {Mortier},
  {Newman}, {Plavchan}, {Quirrenbach}, {Reiners}, {Robertson}, {Roy}, {Schwab},
  {Seifahrt}, {Szentgyorgyi}, {Terrien}, {Teske}, {Thompson}, \&
  {Vasisht}}]{Crass2021}
{Crass}, J., {Gaudi}, B.~S., {Leifer}, S., {et~al.} 2021, arXiv e-prints,
  arXiv:2107.14291, \dodoi{10.48550/arXiv.2107.14291}

\bibitem[{{Dravins}(2008)}]{Dravins2008}
{Dravins}, D. 2008, \aap, 492, 199, \dodoi{10.1051/0004-6361:200810481}

\bibitem[{{Dumusque}(2018)}]{Dumusque2018}
{Dumusque}, X. 2018, \aap, 620, A47, \dodoi{10.1051/0004-6361/201833795}

\bibitem[{{Dumusque} {et~al.}(2014){Dumusque}, {Boisse}, \&
  {Santos}}]{Dumusque2014}
{Dumusque}, X., {Boisse}, I., \& {Santos}, N.~C. 2014, \apj, 796, 132,
  \dodoi{10.1088/0004-637X/796/2/132}

\bibitem[{{Dumusque} {et~al.}(2015){Dumusque}, {Glenday}, {Phillips},
  {Buchschacher}, {Collier Cameron}, {Cecconi}, {Charbonneau}, {Cosentino},
  {Ghedina}, {Latham}, {Li}, {Lodi}, {Lovis}, {Molinari}, {Pepe}, {Udry},
  {Sasselov}, {Szentgyorgyi}, \& {Walsworth}}]{Dumusque2015}
{Dumusque}, X., {Glenday}, A., {Phillips}, D.~F., {et~al.} 2015, \apjl, 814,
  L21, \dodoi{10.1088/2041-8205/814/2/L21}

\bibitem[{{Ervin} {et~al.}(2022){Ervin}, {Halverson}, {Burrows}, {Murphy},
  {Roy}, {Haywood}, {Rescigno}, {Bender}, {Lin}, {Burt}, \&
  {Mahadevan}}]{Ervin2022}
{Ervin}, T., {Halverson}, S., {Burrows}, A., {et~al.} 2022, \aj, 163, 272,
  \dodoi{10.3847/1538-3881/ac67e6}

\bibitem[{{Evershed}(1909)}]{Evershed1909}
{Evershed}, J. 1909, \mnras, 69, 454, \dodoi{10.1093/mnras/69.5.454}

\bibitem[{{Fossum} \& {Carlsson}(2005)}]{Fossum2005}
{Fossum}, A., \& {Carlsson}, M. 2005, \apj, 625, 556, \dodoi{10.1086/429614}

\bibitem[{{Franz} \& {Schlichenmaier}(2009)}]{Franz2009}
{Franz}, M., \& {Schlichenmaier}, R. 2009, \aap, 508, 1453,
  \dodoi{10.1051/0004-6361/200913074}

\bibitem[{{Gizon} {et~al.}(2010){Gizon}, {Birch}, \& {Spruit}}]{Gizon2010}
{Gizon}, L., {Birch}, A.~C., \& {Spruit}, H.~C. 2010, \araa, 48, 289,
  \dodoi{10.1146/annurev-astro-082708-101722}

\bibitem[{{Gizon} {et~al.}(2001){Gizon}, {Duvall}, \& {Larsen}}]{Gizon2001}
{Gizon}, L., {Duvall}, T.~L., J., \& {Larsen}, R.~M. 2001, in Recent Insights
  into the Physics of the Sun and Heliosphere: Highlights from SOHO and Other
  Space Missions, ed. P.~{Brekke}, B.~{Fleck}, \& J.~B. {Gurman}, Vol. 203, 189

\bibitem[{{Gonz{\'a}lez Hern{\'a}ndez} {et~al.}(2020){Gonz{\'a}lez
  Hern{\'a}ndez}, {Rebolo}, {Pasquini}, {Lo Curto}, {Molaro}, {Caffau},
  {Ludwig}, {Steffen}, {Esposito}, {Su{\'a}rez Mascare{\~n}o},
  {Toledo-Padr{\'o}n}, {Probst}, {H{\"a}nsch}, {Holzwarth}, {Manescau},
  {Steinmetz}, {Udem}, \& {Wilken}}]{GonzalezHernandez2020}
{Gonz{\'a}lez Hern{\'a}ndez}, J.~I., {Rebolo}, R., {Pasquini}, L., {et~al.}
  2020, \aap, 643, A146, \dodoi{10.1051/0004-6361/202038937}

\bibitem[{{Gray}(2008)}]{Gray2008}
{Gray}, D.~F. 2008, {The Observation and Analysis of Stellar Photospheres}

\bibitem[{Harris {et~al.}(2020)Harris, Millman, van~der Walt, Gommers,
  Virtanen, Cournapeau, Wieser, Taylor, Berg, Smith, Kern, Picus, Hoyer, van
  Kerkwijk, Brett, Haldane, del R{\'{i}}o, Wiebe, Peterson,
  G{\'{e}}rard-Marchant, Sheppard, Reddy, Weckesser, Abbasi, Gohlke, \&
  Oliphant}]{harris2020array}
Harris, C.~R., Millman, K.~J., van~der Walt, S.~J., {et~al.} 2020, Nature, 585,
  357, \dodoi{10.1038/s41586-020-2649-2}

\bibitem[{{Haywood} {et~al.}(2016){Haywood}, {Collier Cameron}, {Unruh},
  {Lovis}, {Lanza}, {Llama}, {Deleuil}, {Fares}, {Gillon}, {Moutou}, {Pepe},
  {Pollacco}, {Queloz}, \& {S{\'e}gransan}}]{Haywood2016}
{Haywood}, R.~D., {Collier Cameron}, A., {Unruh}, Y.~C., {et~al.} 2016, \mnras,
  457, 3637, \dodoi{10.1093/mnras/stw187}

\bibitem[{{Haywood} {et~al.}(2022){Haywood}, {Milbourne}, {Saar}, {Mortier},
  {Phillips}, {Charbonneau}, {Cameron}, {Cegla}, {Meunier}, \&
  {}}]{Haywood2022}
{Haywood}, R.~D., {Milbourne}, T.~W., {Saar}, S.~H., {et~al.} 2022, \apj, 935,
  6, \dodoi{10.3847/1538-4357/ac7c12}

\bibitem[{Hunter(2007)}]{Hunter2007}
Hunter, J.~D. 2007, Computing in Science \& Engineering, 9, 90,
  \dodoi{10.1109/MCSE.2007.55}

\bibitem[{{Jurgenson} {et~al.}(2016){Jurgenson}, {Fischer}, {McCracken},
  {Sawyer}, {Szymkowiak}, {Davis}, {Muller}, \& {Santoro}}]{Jurgenson2016}
{Jurgenson}, C., {Fischer}, D., {McCracken}, T., {et~al.} 2016, in Society of
  Photo-Optical Instrumentation Engineers (SPIE) Conference Series, Vol. 9908,
  Ground-based and Airborne Instrumentation for Astronomy VI, ed. C.~J.
  {Evans}, L.~{Simard}, \& H.~{Takami}, 99086T, \dodoi{10.1117/12.2233002}

\bibitem[{{Kashyap} \& {Hanasoge}(2021)}]{Kashyap2021}
{Kashyap}, S.~G., \& {Hanasoge}, S.~M. 2021, \apj, 916, 87,
  \dodoi{10.3847/1538-4357/ac05bc}

\bibitem[{{Keller} {et~al.}(2004){Keller}, {Sch{\"u}ssler}, {V{\"o}gler}, \&
  {Zakharov}}]{Keller2004}
{Keller}, C.~U., {Sch{\"u}ssler}, M., {V{\"o}gler}, A., \& {Zakharov}, V. 2004,
  \apjl, 607, L59, \dodoi{10.1086/421553}

\bibitem[{{Kopal}(1950)}]{Kopal1950}
{Kopal}, Z. 1950, Harvard College Observatory Circular, 454, 1

\bibitem[{{Lemen} {et~al.}(2012){Lemen}, {Title}, {Akin}, {Boerner}, {Chou},
  {Drake}, {Duncan}, {Edwards}, {Friedlaender}, {Heyman}, {Hurlburt}, {Katz},
  {Kushner}, {Levay}, {Lindgren}, {Mathur}, {McFeaters}, {Mitchell}, {Rehse},
  {Schrijver}, {Springer}, {Stern}, {Tarbell}, {Wuelser}, {Wolfson}, {Yanari},
  {Bookbinder}, {Cheimets}, {Caldwell}, {Deluca}, {Gates}, {Golub}, {Park},
  {Podgorski}, {Bush}, {Scherrer}, {Gummin}, {Smith}, {Auker}, {Jerram},
  {Pool}, {Soufli}, {Windt}, {Beardsley}, {Clapp}, {Lang}, \&
  {Waltham}}]{Lemen2012}
{Lemen}, J.~R., {Title}, A.~M., {Akin}, D.~J., {et~al.} 2012, \solphys, 275,
  17, \dodoi{10.1007/s11207-011-9776-8}

\bibitem[{{Lin} {et~al.}(2022){Lin}, {Monson}, {Mahadevan}, {Ninan},
  {Halverson}, {Nitroy}, {Bender}, {Logsdon}, {Kanodia}, {Terrien}, {Roy},
  {Luhn}, {Gupta}, {Ford}, {Hearty}, {Laher}, {Hunting}, {McBride}, {Salazar
  Rivera}, {Rajagopal}, {Wolf}, {Robertson}, {Wright}, {Blake}, {Ca{\~n}as},
  {Lubar}, {McElwain}, {Ramsey}, {Schwab}, \& {Stefansson}}]{Lin2022}
{Lin}, A. S.~J., {Monson}, A., {Mahadevan}, S., {et~al.} 2022, \aj, 163, 184,
  \dodoi{10.3847/1538-3881/ac5622}

\bibitem[{{Lites} {et~al.}(1993){Lites}, {Elmore}, {Seagraves}, \&
  {Skumanich}}]{Lites1993}
{Lites}, B.~W., {Elmore}, D.~F., {Seagraves}, P., \& {Skumanich}, A.~P. 1993,
  \apj, 418, 928, \dodoi{10.1086/173450}

\bibitem[{{Llama} {et~al.}(2022){Llama}, {Brewer}, {Zhao}, {Fischer}, \&
  {Szymkowiak}}]{Llama2022}
{Llama}, J., {Brewer}, J., {Zhao}, L., {Fischer}, D., \& {Szymkowiak}, A. 2022,
  in American Astronomical Society Meeting Abstracts, Vol.~54, American
  Astronomical Society Meeting \#240, 401.04

\bibitem[{{L{\"o}hner-B{\"o}ttcher}
  {et~al.}(2018{\natexlab{a}}){L{\"o}hner-B{\"o}ttcher}, {Schmidt},
  {Schlichenmaier}, {Doerr}, {Steinmetz}, \&
  {Holzwarth}}]{Lohner-Bottcher2018a}
{L{\"o}hner-B{\"o}ttcher}, J., {Schmidt}, W., {Schlichenmaier}, R., {et~al.}
  2018{\natexlab{a}}, \aap, 617, A19, \dodoi{10.1051/0004-6361/201832886}

\bibitem[{{L{\"o}hner-B{\"o}ttcher} {et~al.}(2019){L{\"o}hner-B{\"o}ttcher},
  {Schmidt}, {Schlichenmaier}, {Steinmetz}, \&
  {Holzwarth}}]{Lohner-Bottcher2019}
{L{\"o}hner-B{\"o}ttcher}, J., {Schmidt}, W., {Schlichenmaier}, R.,
  {Steinmetz}, T., \& {Holzwarth}, R. 2019, \aap, 624, A57,
  \dodoi{10.1051/0004-6361/201834925}

\bibitem[{{L{\"o}hner-B{\"o}ttcher}
  {et~al.}(2018{\natexlab{b}}){L{\"o}hner-B{\"o}ttcher}, {Schmidt}, {Stief},
  {Steinmetz}, \& {Holzwarth}}]{Lohner-Bottcher2018b}
{L{\"o}hner-B{\"o}ttcher}, J., {Schmidt}, W., {Stief}, F., {Steinmetz}, T., \&
  {Holzwarth}, R. 2018{\natexlab{b}}, \aap, 611, A4,
  \dodoi{10.1051/0004-6361/201732107}

\bibitem[{McKinney {et~al.}(2010)}]{mckinney2010data}
McKinney, W., {et~al.} 2010, in Proceedings of the 9th Python in Science
  Conference, Vol. 445, Austin, TX, 51--56

\bibitem[{{Meunier} {et~al.}(2010){Meunier}, {Lagrange}, \&
  {Desort}}]{Meunier2010a}
{Meunier}, N., {Lagrange}, A.~M., \& {Desort}, M. 2010, \aap, 519, A66,
  \dodoi{10.1051/0004-6361/201014199}

\bibitem[{{Milbourne} {et~al.}(2019){Milbourne}, {Haywood}, {Phillips}, {Saar},
  {Cegla}, {Cameron}, {Costes}, {Dumusque}, {Langellier}, {Latham},
  {Maldonado}, {Malavolta}, {Mortier}, {Palumbo}, {Thompson}, {Watson},
  {Bouchy}, {Buchschacher}, {Cecconi}, {Charbonneau}, {Cosentino}, {Ghedina},
  {Glenday}, {Gonzalez}, {Li}, {Lodi}, {L{\'o}pez-Morales}, {Lovis}, {Mayor},
  {Micela}, {Molinari}, {Pepe}, {Piotto}, {Rice}, {Sasselov}, {S{\'e}gransan},
  {Sozzetti}, {Szentgyorgyi}, {Udry}, \& {Walsworth}}]{Milbourne2019}
{Milbourne}, T.~W., {Haywood}, R.~D., {Phillips}, D.~F., {et~al.} 2019, \apj,
  874, 107, \dodoi{10.3847/1538-4357/ab064a}

\bibitem[{{National Academies of Sciences, Engineering, and
  Medicine}(2018)}]{ESS}
{National Academies of Sciences, Engineering, and Medicine}, . 2018, {Exoplanet
  Science Strategy}, \dodoi{10.17226/25187}

\bibitem[{{National Academies of Sciences, Engineering, and
  Medicine}(2021)}]{Decadal}
{National Academies of Sciences, Engineering, and Medicine}. 2021, {Pathways to
  Discovery in Astronomy and Astrophysics for the 2020s},
  \dodoi{10.17226/26141}

\bibitem[{{Palumbo} {et~al.}(2022){Palumbo}, {Ford}, {Wright}, {Mahadevan},
  {Wise}, \& {L{\"o}hner-B{\"o}ttcher}}]{Palumbo2022}
{Palumbo}, Michael~L., I., {Ford}, E.~B., {Wright}, J.~T., {et~al.} 2022, \aj,
  163, 11, \dodoi{10.3847/1538-3881/ac32c2}

\bibitem[{Palumbo {et~al.}(2023{\natexlab{a}})Palumbo, Saar, \&
  Haywood}]{palumbo_iii_michael_louis_2023_8273650}
Palumbo, M.~L., Saar, S.~H., \& Haywood, R.~D. 2023{\natexlab{a}}, SDO CLV
  Pipeline Output, v2,  Zenodo, \dodoi{10.5281/zenodo.10966717}

\bibitem[{Palumbo {et~al.}(2023{\natexlab{b}})Palumbo, Saar, \&
  Raphaëlle~D.}]{palumbo_michael_louis_2023_8273623}
Palumbo, M.~L., Saar, S.~H., \& Raphaëlle~D., H. 2023{\natexlab{b}},
  sdo-clv-pipeline, v0.2.0,  Zenodo, \dodoi{10.5281/zenodo.10966689}

\bibitem[{{Pepe} {et~al.}(2002){Pepe}, {Mayor}, {Galland}, {Naef}, {Queloz},
  {Santos}, {Udry}, \& {Burnet}}]{Pepe2002}
{Pepe}, F., {Mayor}, M., {Galland}, F., {et~al.} 2002, \aap, 388, 632,
  \dodoi{10.1051/0004-6361:20020433}

\bibitem[{{Pepe} {et~al.}(2000){Pepe}, {Mayor}, {Delabre}, {Kohler}, {Lacroix},
  {Queloz}, {Udry}, {Benz}, {Bertaux}, \& {Sivan}}]{Pepe2000}
{Pepe}, F., {Mayor}, M., {Delabre}, B., {et~al.} 2000, in Society of
  Photo-Optical Instrumentation Engineers (SPIE) Conference Series, Vol. 4008,
  Optical and IR Telescope Instrumentation and Detectors, ed. M.~{Iye} \& A.~F.
  {Moorwood}, 582--592, \dodoi{10.1117/12.395516}

\bibitem[{{Pepe} {et~al.}(2021){Pepe}, {Cristiani}, {Rebolo}, {Santos},
  {Dekker}, {Cabral}, {Di Marcantonio}, {Figueira}, {Lo Curto}, {Lovis},
  {Mayor}, {M{\'e}gevand}, {Molaro}, {Riva}, {Zapatero Osorio}, {Amate},
  {Manescau}, {Pasquini}, {Zerbi}, {Adibekyan}, {Abreu}, {Affolter}, {Alibert},
  {Aliverti}, {Allart}, {Allende Prieto}, {{\'A}lvarez}, {Alves}, {Avila},
  {Baldini}, {Bandy}, {Barros}, {Benz}, {Bianco}, {Borsa}, {Bourrier},
  {Bouchy}, {Broeg}, {Calderone}, {Cirami}, {Coelho}, {Conconi}, {Coretti},
  {Cumani}, {Cupani}, {D'Odorico}, {Damasso}, {Deiries}, {Delabre},
  {Demangeon}, {Dumusque}, {Ehrenreich}, {Faria}, {Fragoso}, {Genolet},
  {Genoni}, {G{\'e}nova Santos}, {Gonz{\'a}lez Hern{\'a}ndez}, {Hughes},
  {Iwert}, {Kerber}, {Knudstrup}, {Landoni}, {Lavie}, {Lillo-Box}, {Lizon},
  {Maire}, {Martins}, {Mehner}, {Micela}, {Modigliani}, {Monteiro}, {Monteiro},
  {Moschetti}, {Murphy}, {Nunes}, {Oggioni}, {Oliveira}, {Oshagh}, {Pall{\'e}},
  {Pariani}, {Poretti}, {Rasilla}, {Rebord{\~a}o}, {Redaelli}, {Santana
  Tschudi}, {Santin}, {Santos}, {S{\'e}gransan}, {Schmidt}, {Segovia},
  {Sosnowska}, {Sozzetti}, {Sousa}, {Span{\`o}}, {Su{\'a}rez Mascare{\~n}o},
  {Tabernero}, {Tenegi}, {Udry}, \& {Zanutta}}]{Pepe2021}
{Pepe}, F., {Cristiani}, S., {Rebolo}, R., {et~al.} 2021, \aap, 645, A96,
  \dodoi{10.1051/0004-6361/202038306}

\bibitem[{{Petersburg} {et~al.}(2020){Petersburg}, {Ong}, {Zhao}, {Blackman},
  {Brewer}, {Buchhave}, {Cabot}, {Davis}, {Jurgenson}, {Leet}, {McCracken},
  {Sawyer}, {Sharov}, {Tronsgaard}, {Szymkowiak}, \&
  {Fischer}}]{Petersburg2020}
{Petersburg}, R.~R., {Ong}, J.~M.~J., {Zhao}, L.~L., {et~al.} 2020, \aj, 159,
  187, \dodoi{10.3847/1538-3881/ab7e31}

\bibitem[{{Reiners}(2009)}]{Reiners2009}
{Reiners}, A. 2009, \aap, 498, 853, \dodoi{10.1051/0004-6361/200810257}

\bibitem[{{Reiners} {et~al.}(2016){Reiners}, {Mrotzek}, {Lemke}, {Hinrichs}, \&
  {Reinsch}}]{Reiners2016}
{Reiners}, A., {Mrotzek}, N., {Lemke}, U., {Hinrichs}, J., \& {Reinsch}, K.
  2016, \aap, 587, A65, \dodoi{10.1051/0004-6361/201527530}

\bibitem[{{Riethm{\"u}ller} {et~al.}(2008){Riethm{\"u}ller}, {Solanki},
  {Zakharov}, \& {Gandorfer}}]{Riethmuller+2008}
{Riethm{\"u}ller}, T.~L., {Solanki}, S.~K., {Zakharov}, V., \& {Gandorfer}, A.
  2008, \aap, 492, 233, \dodoi{10.1051/0004-6361:200810701}

\bibitem[{{Rimmele} \& {Marino}(2006)}]{Rimmele2006}
{Rimmele}, T., \& {Marino}, J. 2006, \apj, 646, 593, \dodoi{10.1086/504794}

\bibitem[{{Rutten}(2003)}]{Rutten2003}
{Rutten}, R.~J. 2003, {Radiative Transfer in Stellar Atmospheres}

\bibitem[{{Saar} {et~al.}(2018){Saar}, {Palumbo}, {Haywood}, \&
  {Dupree}}]{Saar2018}
{Saar}, S.~H., {Palumbo}, Michael~L., I., {Haywood}, R.~D., \& {Dupree}, A.~K.
  2018, in 20th Cambridge Workshop on Cool Stars, Stellar Systems and the Sun,
  Cambridge Workshop on Cool Stars, Stellar Systems, and the Sun, 86,
  \dodoi{10.5281/zenodo.1490103}

\bibitem[{{Scherrer} {et~al.}(2012){Scherrer}, {Schou}, {Bush}, {Kosovichev},
  {Bogart}, {Hoeksema}, {Liu}, {Duvall}, {Zhao}, {Title}, {Schrijver},
  {Tarbell}, \& {Tomczyk}}]{Scherrer2012}
{Scherrer}, P.~H., {Schou}, J., {Bush}, R.~I., {et~al.} 2012, \solphys, 275,
  207, \dodoi{10.1007/s11207-011-9834-2}

\bibitem[{{Schlichenmaier} \& {Schmidt}(1999)}]{Schlichenmaier1999}
{Schlichenmaier}, R., \& {Schmidt}, W. 1999, \aap, 349, L37

\bibitem[{{Schwab} {et~al.}(2016){Schwab}, {Rakich}, {Gong}, {Mahadevan},
  {Halverson}, {Roy}, {Terrien}, {Robertson}, {Hearty}, {Levi}, {Monson},
  {Wright}, {McElwain}, {Bender}, {Blake}, {St{\"u}rmer}, {Gurevich},
  {Chakraborty}, \& {Ramsey}}]{Schwab2016}
{Schwab}, C., {Rakich}, A., {Gong}, Q., {et~al.} 2016, in Society of
  Photo-Optical Instrumentation Engineers (SPIE) Conference Series, Vol. 9908,
  Ground-based and Airborne Instrumentation for Astronomy VI, ed. C.~J.
  {Evans}, L.~{Simard}, \& H.~{Takami}, 99087H, \dodoi{10.1117/12.2234411}

\bibitem[{{SILSO World Data Center}(2012-2015)}]{sidc}
{SILSO World Data Center}. 2012-2015, International Sunspot Number Monthly
  Bulletin and online catalogue

\bibitem[{{Solanki}(2003)}]{Solanki2003}
{Solanki}, S.~K. 2003, \aapr, 11, 153, \dodoi{10.1007/s00159-003-0018-4}

\bibitem[{{Spruit}(1976)}]{Spruit1976}
{Spruit}, H.~C. 1976, \solphys, 50, 269, \dodoi{10.1007/BF00155292}

\bibitem[{{Stief} {et~al.}(2019){Stief}, {L{\"o}hner-B{\"o}ttcher}, {Schmidt},
  {Steinmetz}, \& {Holzwarth}}]{Stief2019}
{Stief}, F., {L{\"o}hner-B{\"o}ttcher}, J., {Schmidt}, W., {Steinmetz}, T., \&
  {Holzwarth}, R. 2019, \aap, 622, A34, \dodoi{10.1051/0004-6361/201834538}

\bibitem[{{Sulis} {et~al.}(2023){Sulis}, {Lendl}, {Cegla}, {Rodr{\'\i}guez
  D{\'\i}az}, {Bigot}, {Van Grootel}, {Bekkelien}, {Cameron}, {Maxted},
  {Simon}, {Lovis}, {Scandariato}, {Bruno}, {Nardiello}, {Bonfanti},
  {Fridlund}, {Persson}, {Salmon}, {Sousa}, {Wilson}, {Krenn}, {Hoyer},
  {Santerne}, {Ehrenreich}, {Alibert}, {Alonso}, {Anglada}, {B{\'a}rczy},
  {Barrado y Navascues}, {Barros}, {Baumjohann}, {Beck}, {Beck}, {Benz},
  {Billot}, {Bonfils}, {Borsato}, {Brandeker}, {Broeg}, {Cabrera}, {Charnoz},
  {Corral van Damme}, {Csizmadia}, {Davies}, {Deleuil}, {Deline}, {Delrez},
  {Demangeon}, {Demory}, {Erikson}, {Fortier}, {Fossati}, {Gandolfi}, {Gillon},
  {G{\"u}del}, {Heng}, {Isaak}, {Kiss}, {Laskar}, {Lecavelier des Etangs},
  {Magrin}, {Munari}, {Nascimbeni}, {Olofsson}, {Ottensamer}, {Pagano},
  {Pall{\'e}}, {Peter}, {Piotto}, {Pollacco}, {Queloz}, {Ragazzoni}, {Rando},
  {Rauer}, {Ribas}, {Rieder}, {Santos}, {S{\'e}gransan}, {Smith},
  {Steinberger}, {Steller}, {Szab{\'o}}, {Thomas}, {Udry}, {Walton}, \&
  {Wolter}}]{Sulis2023}
{Sulis}, S., {Lendl}, M., {Cegla}, H.~M., {et~al.} 2023, \aap, 670, A24,
  \dodoi{10.1051/0004-6361/202244223}

\bibitem[{{SunPy Community} {et~al.}(2020){SunPy Community}, {Barnes}, {Bobra},
  {Christe}, {Freij}, {Hayes}, {Ireland}, {Mumford}, {Perez-Suarez}, {Ryan},
  {Shih}, {Chanda}, {Glogowski}, {Hewett}, {Hughitt}, {Hill}, {Hiware},
  {Inglis}, {Kirk}, {Konge}, {Mason}, {Maloney}, {Murray}, {Panda}, {Park},
  {Pereira}, {Reardon}, {Savage}, {Sip{\H{o}}cz}, {Stansby}, {Jain}, {Taylor},
  {Yadav}, {Rajul}, \& {Dang}}]{SunPyCommunity2020}
{SunPy Community}, {Barnes}, W.~T., {Bobra}, M.~G., {et~al.} 2020, \apj, 890,
  68, \dodoi{10.3847/1538-4357/ab4f7a}

\bibitem[{Tange(2021)}]{tange_2021_5523272}
Tange, O. 2021, GNU Parallel 20210922 ('Vindelev'),  Zenodo,
  \dodoi{10.5281/zenodo.5523272}

\bibitem[{{Thompson}(2006)}]{Thompson2006}
{Thompson}, W.~T. 2006, \aap, 449, 791, \dodoi{10.1051/0004-6361:20054262}

\bibitem[{{Tiwari} {et~al.}(2013){Tiwari}, {van Noort}, {Lagg}, \&
  {Solanki}}]{Tiwari2013}
{Tiwari}, S.~K., {van Noort}, M., {Lagg}, A., \& {Solanki}, S.~K. 2013, \aap,
  557, A25, \dodoi{10.1051/0004-6361/201321391}

\bibitem[{Van~der Walt {et~al.}(2014)Van~der Walt, Sch{\"o}nberger,
  Nunez-Iglesias, Boulogne, Warner, Yager, Gouillart, \& Yu}]{van2014scikit}
Van~der Walt, S., Sch{\"o}nberger, J.~L., Nunez-Iglesias, J., {et~al.} 2014,
  PeerJ, 2, e453

\bibitem[{Virtanen {et~al.}(2020)Virtanen, Gommers, Oliphant, Haberland, Reddy,
  Cournapeau, Burovski, Peterson, Weckesser, Bright, {van der Walt}, Brett,
  Wilson, Millman, Mayorov, Nelson, Jones, Kern, Larson, Carey, Polat, Feng,
  Moore, {VanderPlas}, Laxalde, Perktold, Cimrman, Henriksen, Quintero, Harris,
  Archibald, Ribeiro, Pedregosa, {van Mulbregt}, \& {SciPy 1.0
  Contributors}}]{2020SciPy-NMeth}
Virtanen, P., Gommers, R., Oliphant, T.~E., {et~al.} 2020, Nature Methods, 17,
  261, \dodoi{10.1038/s41592-019-0686-2}

\bibitem[{{Westendorp Plaza} {et~al.}(1997){Westendorp Plaza}, {del Toro
  Iniesta}, {Ruiz Cobo}, {Martinez Pillet}, {Lites}, \&
  {Skumanich}}]{WestendorpPlaza1997}
{Westendorp Plaza}, C., {del Toro Iniesta}, J.~C., {Ruiz Cobo}, B., {et~al.}
  1997, \nat, 389, 47, \dodoi{10.1038/37933}

\bibitem[{{Wilson} \& {Maskelyne}(1774)}]{Wilson1774}
{Wilson}, A., \& {Maskelyne}, N. 1774, Philosophical Transactions of the Royal
  Society of London Series I, 64, 1

\bibitem[{{Wright} \& {Eastman}(2014)}]{Wright2014}
{Wright}, J.~T., \& {Eastman}, J.~D. 2014, \pasp, 126, 838,
  \dodoi{10.1086/678541}

\bibitem[{{Yeo} \& {Krivova}(2019)}]{Yeo2019}
{Yeo}, K.~L., \& {Krivova}, N.~A. 2019, \aap, 624, A135,
  \dodoi{10.1051/0004-6361/201935123}

\bibitem[{{Yeo} {et~al.}(2013){Yeo}, {Solanki}, \& {Krivova}}]{Yeo2013}
{Yeo}, K.~L., {Solanki}, S.~K., \& {Krivova}, N.~A. 2013, \aap, 550, A95,
  \dodoi{10.1051/0004-6361/201220682}

\bibitem[{{Zhao} \& {Dumusque}(2023)}]{Zhao2023}
{Zhao}, Y., \& {Dumusque}, X. 2023, \aap, 671, A11,
  \dodoi{10.1051/0004-6361/202244568}

\end{thebibliography}
\bibliographystyle{aasjournal}

\end{document}